\documentclass[aps,pre,nofootinbib, amsfonts,superscriptaddress,longbibliography,showkeys,notitlepage]{revtex4-1}

\usepackage{graphicx}
\usepackage{tabularx}
\usepackage{braket}
\usepackage{amssymb}
\usepackage{color}
\usepackage{placeins}
           
\begin{document}
%opening
\title{Dynamics of heuristics selection for cooperative behavior}

\author{Felipe Maciel Cardoso}
\affiliation{Institute for Biocomputation and Physics of Complex Systems (BIFI), University of Zaragoza, Spain}
\author{Carlos Gracia-L\'azaro}
\affiliation{Institute for Biocomputation and Physics of Complex Systems (BIFI), University of Zaragoza, Spain}
\affiliation{Department of Theoretical Physics, Faculty of Sciences, University of Zaragoza, Spain}
\author{Yamir Moreno}
\affiliation{Institute for Biocomputation and Physics of Complex Systems (BIFI), University of Zaragoza, Spain}
\affiliation{Department of Theoretical Physics, Faculty of Sciences, University of Zaragoza, Spain}
\affiliation{ISI Foundation, Turin, Italy}

\begin{abstract}
Situations involving cooperative behaviour are widespread among animals and humans alike. Game theory and evolutionary dynamics have provided the theoretical and computational grounds to understand what are the mechanisms that allow for such cooperation. Studies in this area usually take into consideration different behavioural strategies and investigate how they can be fixed in the population under evolving rules. However, how those strategies emerged from basic evolutionary mechanisms continues to be not fully understood. To address this issue, here we study the emergence of cooperative strategies through a model of heuristics selection based on evolutionary algorithms. In the proposed model, agents interact with other players according to a heuristic specified by their genetic code and reproduce -- at a longer time scale -- proportionally to their fitness. We show that the system can evolve to cooperative regimes for low mutation rates through heuristics selection while increasing the mutation decreases the level of cooperation. Our analysis of possible strategies shows that reciprocity and punishment are the main ingredients for cooperation to emerge, being conditional cooperation the more frequent strategy. Additionally, we show that if in addition to behavioural rules, genetic relatedness is included, then kinship plays a relevant role. Our results illustrate that our evolutionary heuristics model is a generic and powerful tool to study the evolution of cooperative behaviour. 
\end{abstract}


\renewcommand\tabularxcolumn[1]{m{#1}}% for vertical centering text in X column

\maketitle

\section{Introduction}

Game theory constitutes a powerful framework for the mathematical study of social dilemmas \cite{VonNeumann1944, Myerson1997}. Within this framework, the most representative and widely used game to model cooperation, the Prisoner's Dilemma, has become a paradigm for modelling the evolution of cooperative behaviour \cite{Rapoport1965}. The Prisoner's Dilemma mimics the worst possible scenario for cooperation in which selfishness always provides a higher individual benefit than cooperative behaviour. 
Initial predictions indicated the social optimum would not be reachable by rational selfish individuals if the temptation for defecting ($T$) exceeded the reward for cooperating $R$.
Nonetheless, cooperation is pervasive in human and animal societies \cite{Dugatkin1997, Bourke2011, Bowles2011}, and a vast literature has demonstrated how cooperation can thrive in the presence of an appropriate evolutionary process \cite{Nowak1992, Nowak1992TfT, Nowak2006, Axelrod1987, Lindgren1991, SANTOS2006, Roca2009, Colman2012, Sigmund2010}. 
The possible situations where cooperation might flourish are endless, and we are just beginning to uncover the ingredients behind the complexity observed in real systems \cite{Henrich2005, Foster2005}. Consequently, theoretical studies usually focus on simplifications, such as individuals behaving according to fixed pure strategies \cite{Nowak1992, SANTOS2006} or some arbitrary set of them  \cite{Traulsen2009, Su2016}.  Yet, the reasoning and motivations of humans are more sophisticated and complex than pure strategies and decisions are usually taken factoring in many ingredients, weighting them differently \cite{Tversky1974}. In other words, generally speaking, the selection of strategies takes place in
complex systems wherein imprecise behaviour and the environment are inputs of each other in a perpetual feedback loop \cite{Gintis2011}.

In this line, behavioural economics has shown that humans respond in unexpected ways \cite{Aumann2019a, Thaler2018} and often seem to possess hardwired heuristics while acting in experimental situations \cite{Knafo2008, Bateson2006}. Experiments have also shown that humans automatic responses are modelled by experiences from daily-life, building heuristics or intuitions which tend to favour cooperation \cite{Rand2012, Rand2014a}. Therefore, it is plausible that cooperative societies are sustained by existent heuristics, maintained by norms \cite{Boyd2018, Richerson2016} or biological factors \cite{Hamilton1964, Foster2005, Knafo2008}, that have resulted from a selection dynamics. It is thus imperative to understand how such possible heuristics have evolved, which will allow explaining the ingrained mechanisms behind the behaviour observed in living beings. 

\begin{figure}
	\includegraphics[width=0.9\linewidth]{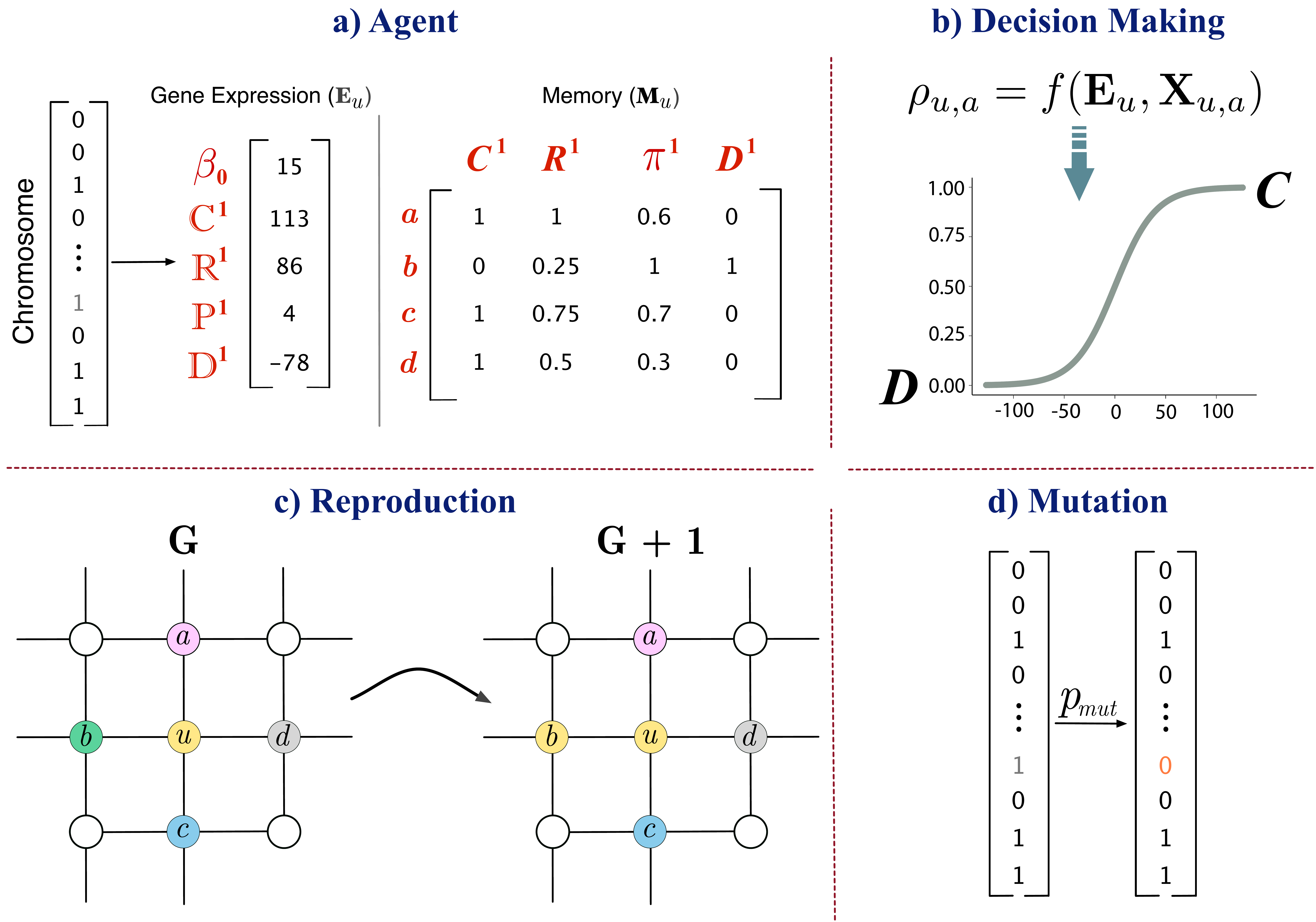} 
	\caption{\textbf{Illustration of the model for memory 1.} 
		\textbf{a)} Agents are composed of a chromosome and memory. Their memories store their experiences with their neighbours, and their chromosomes determine what will be their responses to the variables stored.
		\textbf{b)} Agent $u$ cooperates probabilistically with agent $a$ according to what is coded in its gene and the history of agent $a$.
		\textbf{c)} Reproduction takes place synchronously at the end of a generation ($G$): for each site $u$, a new agent is chosen proportionally to its fitness from the set $\{u \cap N(u)\}$ (coloured nodes), wherein $N(u)$ are $u$'s neighbours. In the example, each colour corresponds to a different chromosome and, at generation $G+1$, the chromosome of agent $u$ happens to have reproduced in sites $u$ and $b$, while its other neighbours by chance maintained the same chromosome.
		\textbf{d)} When an agent reproduces, with probability $p_{mut}$ a bit will be flipped.} 
\label{fig:diagram}
\end{figure}

In this paper, we investigate the evolution of cooperative strategies through an agent-based model of heuristics selection inspired by evolutionary algorithms \cite{Eiben2003}. The ultimate goal is to obtain a description of the evolutionary process that could lead to different strategies. Explicitly, we consider agents composed of a chromosome and memory to store information of other players' previous actions (Fig. \ref{fig:diagram}a). Their actions are responses, according to what is coded in their genes, to other players' history. The strategy space is given, thus, by all the possible genes' combinations. This does not mean that we model behaviours defined by real genomes: decision making, especially in humans, has entangled layers of complexity, and such an approach would be misguided. Rather, we use chromosomes as a tool to model heuristics formed through cultural or biological evolution \cite{Richerson2005, Gintis2011}.

In our framework, the fitness of agents corresponds to the payoff obtained in iterated games, and it determines the agent's reproduction rates. Offspring will inherit its parents' chromosomes while being susceptible to mutation. Note that our approach differs from elementary evolutionary algorithms: they optimize functions in a constant fitness landscape, but in evolutionary games changes in the population imply changes in the fitness landscape \cite{Nowak2004b}, which can be easily seen in any form of the rock-paper-scissors game \cite{Sigmund2010}.

The use of evolutionary algorithms to explore the adaptation of agents is not new \cite{Lindgren1991, Axelrod1987, Miller1996}, and previous works have studied the evolution of automata-like strategies, though aiming at answering specific situations \cite{VanVeelen2012, Binmore1992}. In these studies, the equivalent of a chromosome is a tool to encode an extensive set of memory-based strategies used to understand when cooperation may thrive. Unfortunately, these types of strategies are hardly realistic and do not correspond to the optimal model for understanding the mechanisms behind human or animal responses. A model of heuristics should resemble more closely automatic responses based on intuition and past experiences \cite{Tversky1974}, namely, by considering that intuitive responses are no more than stochastic processes which take as inputs the variables observed by the individual. 

Here, we develop a modelling approach in which agents can evaluate different variables at the same time, thus resembling real situations wherein different factors interact and affect actions. Agents decisions are determined by an activation function taking as input their chromosome and the information to which they have access. Given their theoretical and practical importance, we focus on the evolution of cooperation in social dilemmas. For this case, therefore, we selected a set of variables based on the history of the players with whom they are playing. Nonetheless, our modelling framework is generic, and any arbitrary set of variables can be added or removed according to the question of interest. Our results show that the specified heuristics can evolve to cooperative equilibria for low mutation rates. An analysis of agents chromosome reveals that cooperation endures by reciprocity, indicating that the evolution drives heuristics to reproduce a fundamental mechanism underlying cooperation in nature, especially in humans \cite{West2007a, Clutton-Brock2009a}. In this case, emerging strategies of conditional cooperators dominate, permitting cooperation to prosper. Finally, we provide an extension wherein agents can evaluate their genetic relatedness with others. The population in this scenario evolves to similar equilibria. However, the agents' chromosomes differ significantly from the first model. Kin identification becomes the main mechanism of cooperative heuristics. Nonetheless, agents still need to have a memory of their past actions for cooperation to endure.  

Undoubtedly, varied environmental or perception variables affect the resulting behaviour in humans and other animals. Unfortunately, it is not straightforward to capture which variables guided evolution to the emerged behaviour in each particular scenario. In this line, our proposal provides one generic approach for the modelling of such processes. In particular, the model here presented also contribute some insightful results with the current specifications. Namely, we observe that cooperation can spread spontaneously when memory is available, and that mutation is essential to ensure this outcome. Moreover, although the same behaviour might be observed in distinct populations, the underlying causes might be significantly different, as we observe with our kin and non-kin models. These insights suggest that our method can be a useful tool to uncover the ultimate causes behind the evolution of pro-social behaviour.

\section{The model}

\subsection{Population Dynamics}

We consider a virtual environment inhabited by $n$ haploid agents in a zero population growth condition, each one of them ($u$) containing a chromosome $\textbf{A}_u$ defining the heuristic which will guide its decision. Each agent interacts with each other through links defined by a static contact structure, in which $L$ is the set of edges connecting the two pairs of agents. In real systems, a generation embodies repeated interactions between individuals, and it is known that fast selection fluctuations can suppress cooperation even in the cases in which it is the only rational choice \cite{Roca2009}. In our model, in each generation, there is a finite number of $s=100$ time steps and, therefore, $s|L|$ dyadic interactions take place, \textit{i.e.}, one for each edge at each time step. Thus, at each time step $t$, connected agents $u$ and $v$ interact in a game and obtain the payoffs $\pi_u^t$ and $\pi_v^t$, respectively. 

The generation reaches its end after the $s$ time steps, and each agent $u$ will have accumulated a total payoff of $\Pi_u$, corresponding to its fitness in a strong selection pressure process \cite{Traulsen2007, Roca2009}. 
Agents reproduce by a localized \textit{death-birth} process \cite{Lindgren1994}: at the end of each generation, each node $u$ will be replaced by a node $u'$ in the set $N_2(u)= N(u) \cap \{u\}$, which is composed by the neighbourhood of $u$ ($N(u)$) and $u$ itself (Fig. \ref{fig:diagram}c). 
Node $u'$ is chosen probabilistically according to the fitness $(\Pi_{u'})$ of nodes in $N_2$. Thus, on one hand, the nodes which accumulate more payoff are more likely to be chosen, on the other hand, most adapted agents can reproduce up to sites of distance one. 
Finally, some fluctuations might affect offspring. Specifically, there is a probability $p_{mut}$ of a newborn having a bit flipped in their chromosome (Fig. \ref{fig:diagram}d).

\subsection{Game}

We are interested in the evolution of cooperation in a population of agents facing a social dilemma. Strictly speaking, we want to check if cooperative heuristics are the most adapted in conditions wherein pure strategies equilibria would be of full defection. We consider that at each interaction, agents play a round of a Prisoners' Dilemma (PD) with their neighbours. The PD game is a 2x2 game in which only two actions are available to the players, either cooperate or defect. If two players cooperate, they both get a reward $R$, if one cooperates and the other defects, the cooperator earns $S$ and the defector gets a payoff $T$ (the temptation to defect). Finally, if both defect, both of them obtain $P$. The PD occurs when the elements of the payoff matrix are such that $T>R>P>S$,  which implies that a rational player should defect because, whatever your opponent does, the best (in terms of having larger payoff) is to defect. Henceforth, we consider that the values of each entry are a normalized version of Axelrod's tournament \cite{Axelrod1981} values. Namely: $T=1/\braket{k}; R=0.6/\braket{k}; P=0.2/\braket{k}; S = 0$. As mentioned before, for these values, the prediction is that under a replicator dynamics, the system ends up in full defection \cite{Roca2009}. We also note that small changes in this parametrization would not affect our results, as they are robust for a broad range of the temptation ($T$) parameter (see Section \ref{sec:payoffT} of the Supplementary Material).

\subsection{Agents}

Agents are hardwired, and their heuristics do not change in the course of one generation, which corresponds to their lifetime. Their heuristics are determined by their chromosomes and constitute a stochastic way to evaluate the variables stored in their memory and make a decision on whether to cooperate or not. Agents' memory stores the variables from previous interactions, and we assume their working memory is limited \cite{Milinski1998}. Hence, agents can only store a finite set of variables from the previous $m$ rounds. Specifically, an agent $u$ with the set of neighbours $N(u)$, has stored in its memory $\mathbf{M_u}$ variables for all $v \in N(u)$ and for all $l \in [1, m]$. Therefore, $\mathbf{M_u}$ is a matrix wherein each row contains the values stored for one neighbour, as shown in Fig. \ref{fig:diagram}a. 

\begin{table}
	\begin{tabularx}{0.48\linewidth}{|c|c|X|}
		\hline
		\textbf{Variable} & \textbf{Gene}  & \textbf{Description} \\
		\toprule
		& $\beta^0_u$ &  \textbf{Constant response.} \\
		\hline
		$C_{v,u}^l$ & $\mathbb{C}^l_{u}$ & \textbf{Direct Reciprocity:} 1 if $v$ cooperated with $u$ in round $t-l$, 0 otherwise.   \\
		\hline     
		$R_{v, u}^l$ & $ \mathbb{R}^l_{u}$ & \textbf{Indirect Reciprocity:} Fraction of times agent $v$ cooperates in round $t-l$ with players other than $u$.  \footnote{Typically indirect reciprocity is defined by individuals playing a one-shot game in a large well-mixed population \cite{Nowak2005}. Nonetheless, here $R^t_{u,v}$ does not consider the actions of $v$ with respect to $u$, which should correspond to an analogous effect.} \\
		\hline 
		$\pi_v^l$ & $ \mathbb{P}^l_{u}$ & \textbf{Payoff} obtained by agent $v$ in round $t-l$.  \\
		\hline     
		$D_{v, u}^l$ & $ \mathbb{D}^l_{u}$ & \textbf{Punishment:} 0 if $v$ cooperated with $u$ in round $t-l$, 1 otherwise.  \\
		\hline
	\end{tabularx}
	\caption{\textbf{Memory variables and genes.} Genes determine agents actions, and they are responses to the variables stored in their memory. Here we show the variables considered and their corresponding gene for a previous round $l$. The description indicates how the information is stored and the role of each gene.}\label{tb:vargene}.
\end{table}

The heuristics evaluate each stored variable according to a specific gene in the chromosome.  Therefore, the expression of each gene is a weight given to a variable containing some information influencing agents' decision making. The vector $\mathbf{E}_u$ carries the responses of an agent $u$, i.e., its expressed genes values.  They are given by a two complement representation of the gene bits, therefore, they are integers from -128 to 127\footnote{Hence, when a mutation occurs in a gene, from its expressed value it can be added/subtracted a random power of 2, or have its sign and value changed.}. The vector $E_u$ contains the responses to the variables plus a constant response ($\beta^0$).  Table \ref{tb:vargene} shows the set of variables stored and their corresponding genes. They are a basic set of external characteristics that an elementary agent can observe. Thus, they constitute a reasonable set of variables to be taken into account by a somewhat minimal heuristic. Finally, whether or not an agent will cooperate is determined by the sigmoid function specified by
\begin{equation} \label{eq:decisionf}
\rho_{u, v} = f(\mathbf{E}_u, \mathbf{X}_{u, v}) = \frac{1}{1+e^{-\kappa(\mathbf{E}_u \cdot \mathbf{X}_{u, v})}},
\end{equation}
where $\rho_{u, v}$ corresponds to the probability of agent $u$ cooperating with agent $v$. $\mathbf{X}_{u,v}=(1) \oplus \mathbf{M}_{u,v}$ corresponds to a vector composed by the number 1 in the first position, followed by the memory variables as specified in Table \ref{tb:vargene}. $\kappa$ (henceforth set to 0.05) provides the steepness of the curve, and it is chosen in such a way that if the dot product of both vectors is greater (\textit{resp. lesser}) than 100, the probability should be approximately 1 (\textit{resp. 0}), as illustrated by Fig. \ref{fig:diagram}b.

\section{Results}

\begin{figure}
	\includegraphics[width=0.5\textwidth]{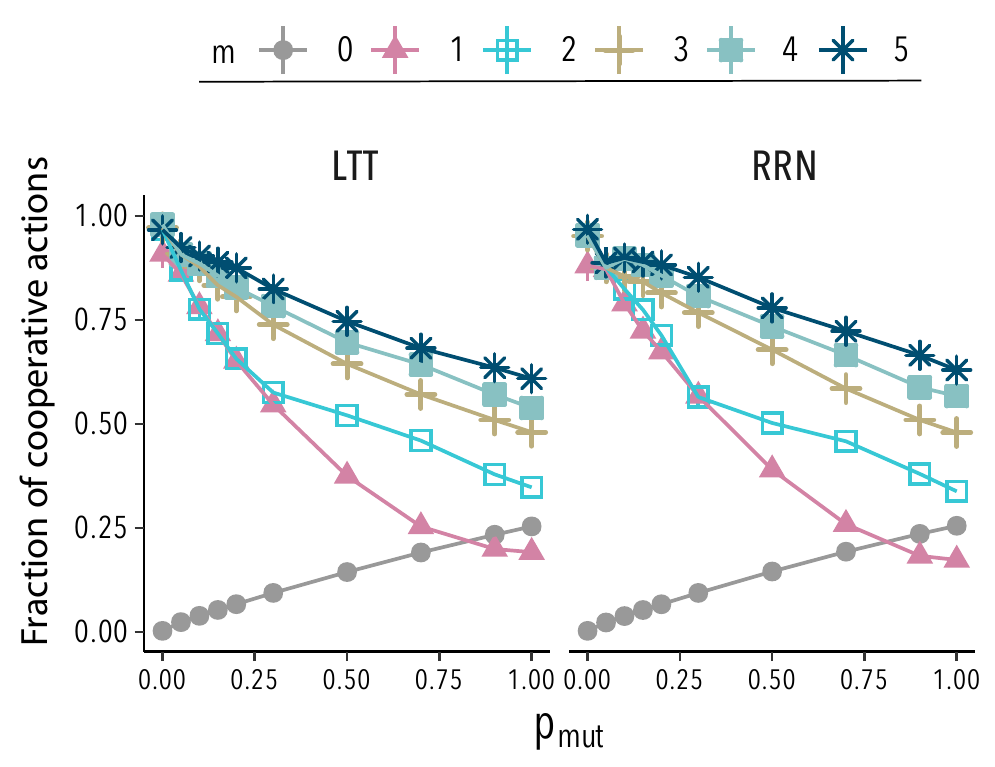}
	\caption{\textbf{Cooperation thrives at low mutation values}. Fraction of cooperative actions at the steady-state according to the mutation probability. Colours and shapes correspond for different memory ($m$) values.  Averages plus .95 confidence interval of 100 realizations are presented for each mutation ($p_{mut}$) value. }  \label{fig:pMperCmemory}
\end{figure}

We ran simulations for populations of 1024 agents connected on a lattice (\textit{LTT}) with a von Neumann neighbourhood and on Random Regular Networks (\textit{RRN}) with the same nodes' degree ($k=4$).  We evolved the model for $5 \cdot 10^5$ generations, each with 100 rounds, for different values of the $p_{mut}$ parameter. Results for memory between 0 and 5 are shown in Fig. \ref{fig:pMperCmemory}. When $m=0$, the agents' chromosome is composed of only the constant response ($\beta^0$) and strategies are reducible to mixed strategies. In this case, when no mutation is available the system quickly goes to full defection (see Supplementary Fig. \ref{fig:SIperC0}), as expected, and mutation increases the possibility of adding cooperative strategies by drift. Conversely, when agents have access to memory, cooperation is predominant in the regime of low mutation. Furthermore, cooperation is larger and more resilient to higher mutation when agents have access to a bigger memory. With more memory, agents should be able to construct more complex heuristics which seem to favour cooperation.

\begin{figure}
	\includegraphics[width=0.5\textwidth]{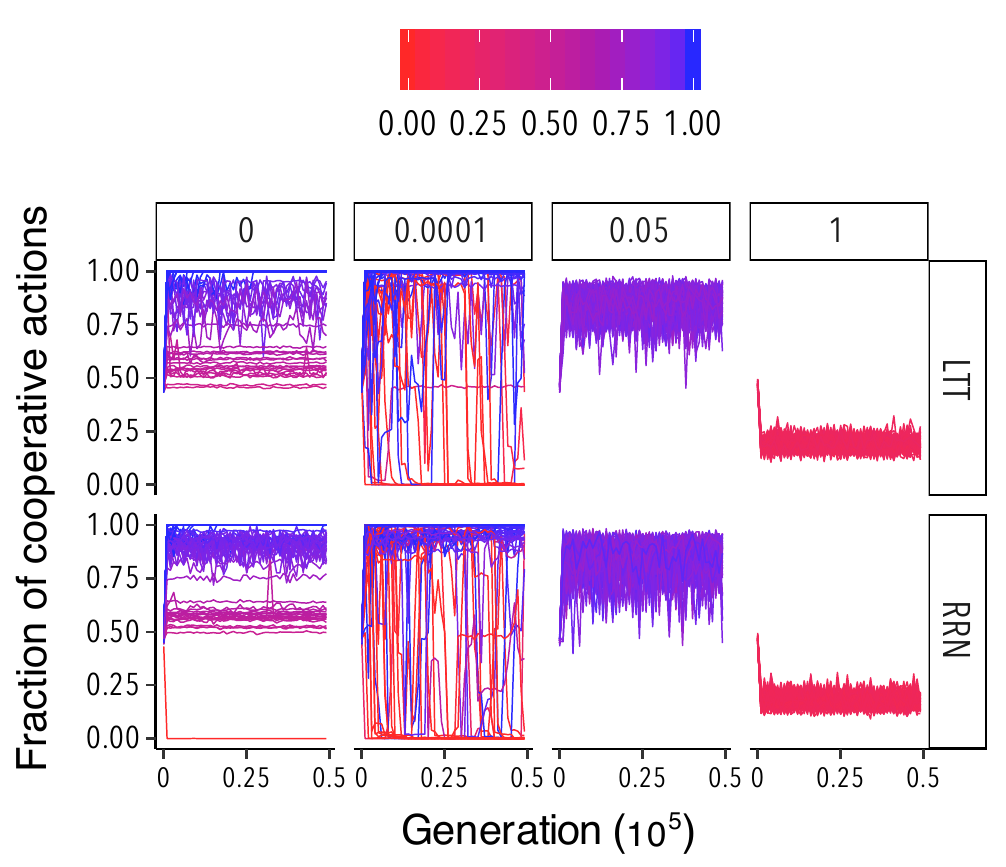}
	\caption{\textbf{Fraction of cooperative actions at the end of each generation}. Columns correspond to different mutation values (0, 0.0001, 0.05, 1) and horizontal panels to different networks (LTT, RRN). Agents have memory $m=1$ and 100 realizations are performed for each mutation value and network. The colours of the lines correspond to the average of the last 1000 generations.}  \label{fig:perC1}
\end{figure}

Figure \ref{fig:perC1} shows time evolution curves of individual realizations for $m=1$ (results for other values of the memory are reported in Section \ref{sec:timeEvo} of the Supplementary Material). When $p_{mut}=0$, the final fraction of cooperative fraction is highly dependent on the initial conditions, reaching a multitude of equilibria, some being fully cooperative and others showing a rather small level of cooperation, specially in the RRN network. In the regime of small mutation rates, fluctuations increase significantly. However, for some small values of the mutation rate, all realizations converge to highly cooperative equilibria, as can be seen when $p_{mut}=0.05$. Note, additionally, that as the probability of mutation increases, the fraction of cooperative actions decreases. For the limiting value $p_{mut} = 1$, every new player is born with a mutation and the system evolves into a negligible average level of cooperation. Interestingly, this is a demonstration that a small noise can foster cooperation in the process of evolution. With more mutation, it gets harder for cooperative strategies to prevail and defection tends to increase, however, a sufficiently small mutation probability will guarantee that the system evolves to a cooperative equilibrium.

\subsection{Heuristics and Strategies}

\begin{figure}
	\includegraphics[width=0.5\textwidth]{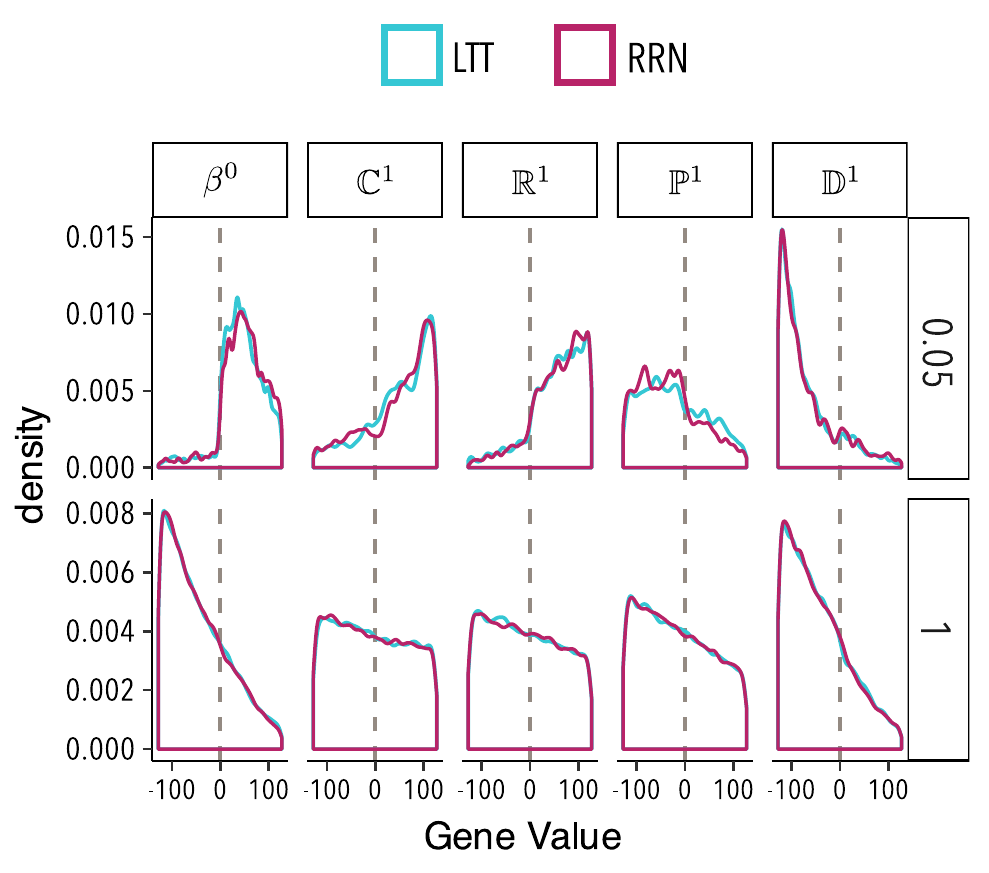}
	\caption{\textbf{\textbf{Distribution of genes' expressed values.}} Densities of genes values for simulations on LTT and RRN graphs for $m=1$. Top panels show distributions for $p_{mut}=0.05$ and bottom panels for $p_{mut}=1$. The vertical dashed line indicates separate regions wherein the marginal probability to cooperate would be smaller (negative gene values) and greater (positive gene value) than 0.5.}  \label{fig:geneDist}
\end{figure}

In this section, we focus on the composition of the populations in the different regimes. It is not straightforward to evaluate how genes and variables interact, hence, it is hard to determine if agents are going to cooperate or not in a specific situation. A first step is to investigate what are the gene values in cooperative and non-cooperative equilibria. Fig. \ref{fig:geneDist} show the distributions of genes for two mutation values: $p_{mut}=0.05$ and $p_{mut}=1$, wherein evolution leads to mostly cooperation and to mostly defection, respectively. Simulations in both LTT and RRN networks yielded similar distributions, indicating the presence of a common evolutionary pattern.

When the majority of the population cooperates ($p_{mut}=0.05$), $\beta^0$, $\mathbb{C}^1$, and  $\mathbb{R}^1$ have a clear right-modality with most of these values being higher than 0. Conversely, $\mathbb{D}^1$ is left-modal with a clear peak at extreme negative values, while $\mathbb{P}^1$ shows a softer trend towards negative values.  This implies that when cooperation thrives, agents have a baseline cooperative response and tend to reciprocate cooperation both directly and indirectly. On the other hand, the agents punish defectors rigorously and have a mild negative response to other agents' payoff, probably as a means to punish defectors, as only defectors can attain the highest payoffs. Interestingly, the distributions of $\beta^0$ indicate that the emerging strategies are willing to cooperate even in a one-shot game (see Supplementary Fig. \ref{fig:SIoneshot}) with an unknown player, albeit this is not the expected behaviour for $m=0$. In the other extreme, for $p_{mut}=1$, defection prevails, and genes values indicate the underpinnings of this trend. All distributions are right-skewed, with  $\beta^0$ and $\mathbb{D}^1$ having a noticeable peak at the lowest possible values. Thus, when mutations are too frequent agents are much more likely to exploit and punish, leading defection to be the default strategy. Too much drift will make it impossible for cooperative heuristics to be selected, and they will vanish in the population.

These last results provide a picture of the genotype space. However, there is still the need to identify which strategies have emerged. When studying evolutionary games, it is always challenging to bridge the gap between the genotype and phenotype spaces \cite{Nowak2004b}. In our model, the profile of agents' actions would correspond to observable phenotypes, yet it is not straightforward to specify a method for heuristics classification. An unsupervised procedure would fall into the problem of how to identify the groups encountered, i.e., how to determine to which known strategies they correspond. Therefore, here we adopted an approach that consisted of classifying agents by looking at what would be their responses to the most basic strategies: a pure defector and a pure cooperator. Namely, we looked at whether agents were likely to cooperate or defect with agents having a history corresponding to each of the two pure strategies. For instance, a full defector $v$ would always have defected with $u$ ($C_{v,u}^1 = 0$, $D_{v, u}^1= 1$), with its other neighbours ($R_{v, u}^1=0$), and would have an expected payoff ($\pi_v^1$) corresponding to these actions. 

\begin{table}
	\centering
	\begin{tabular}{|c|c|c|}
		\hline 
		& \textbf{Pure Cooperator} & \textbf{Pure Defector}   \\ 
		\toprule
		FC & $C$  & $C$  \\ \hline 
		FD & $D$ & $D$ \\ \hline  
		CC & $C$  & $D$   \\ \hline 
		GCC & $C$  & - \\          \hline
		CD & - & $D$ \\  \hline
		Bully & $D$ & $C$ \\  \hline
		Random &  -  & - \\   \hline
		%        Undefined & ? & ? \\  \hline
	\end{tabular} 
	\caption{\textbf{Classification of heuristics according to their responses to the two pure strategies:} pure defector and pure cooperator. We consider that agents cooperate ($C$) or defect ($D$) if their probability to cooperate is greater than $(1 - \sigma)$ or smaller than $\sigma$, respectively.}
	\label{tb:strategies}
\end{table}

The proposed classification is shown in Table \ref{tb:strategies} (see the details in Section \ref{sec:classification} of the SI). We considered strategies analogous to known ones, namely: \textit{Full Cooperator (FC)}, cooperates with both pure cooperators and pure defectors; \textit{Full Defector (FC)}, defects with both; \textit{Conditional Cooperator(CC)}, reciprocates cooperation and defects otherwise;  \textit{Generous Conditional Cooperator(GCC)}, reciprocates cooperation and can cooperate randomly with defectors; \textit{Conditional Defector(CD)}, cooperates randomly with cooperators and always defects with defectors; \textit{Bully}, defects with cooperators, but cooperates with defectors; \textit{Random}, behave randomly with both pure strategies. We labelled agents that could not be classified by this process as \textit{Undefined}. 

\begin{figure}
	\includegraphics[width=0.5\textwidth]{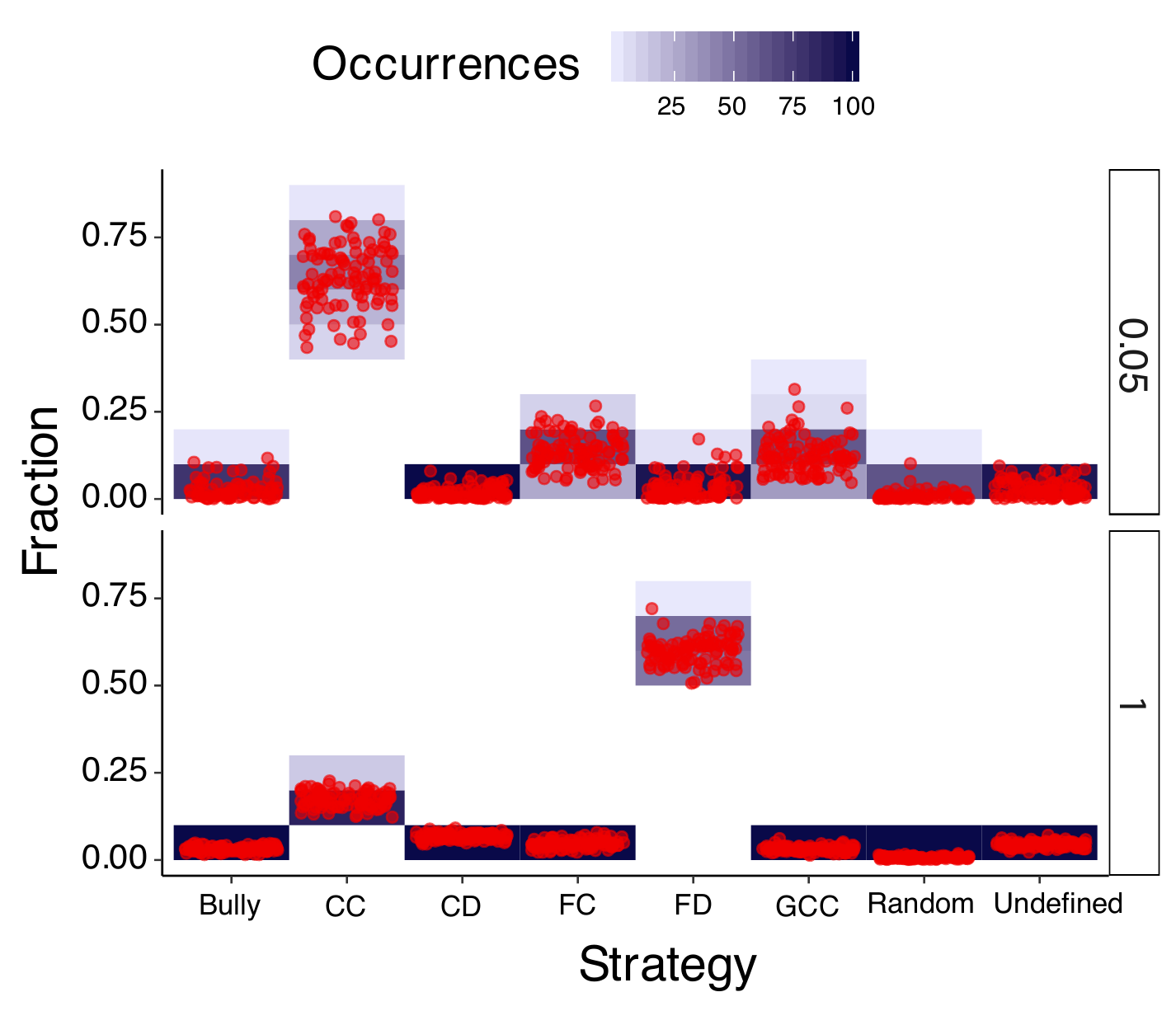}
	\caption{\textbf{\textbf{Emerging Strategies in a lattice.}} Frequency of each strategy in executions in LTT networks for $p_{mut}=0.05$, $\sigma=0.3$ and $m=1$. Each red dot correspond to the fraction of the strategy in a simulation and the histogram of fractions for each strategy is shown vertically, with darkest colours representing a higher number of occurrences. }  \label{fig:strategies}
\end{figure}

In Fig. \ref{fig:strategies}, we show the frequencies of each strategy from simulations of the heuristics selection dynamics on a lattice (a similar pattern is obtained for RRN networks, see the SI, Fig. \ref{fig:SIstrategies}). When the mutation is low ($p_{mut}=0.05$), most of the agents tend to be cooperators or conditional cooperators (mean fraction is 0.9 with a standard deviation of 0.07): CC constitutes most of the strategies, followed by a small fraction of GCC and FC  players. In contrast, when mutation is high ($p_{mut}=1$), FD and CD constitute the majority (mean=0.66, sd=0.038) of agents. However, a minority of CC players can persist (mean=0.17, sd=0.022), which explains the existence of a small fraction of cooperative actions even in this regime.

\section{Exploring kin discrimination: a first extension.}

It is known that cooperative behaviour can emerge and be sustained by factors that do not depend on players history of decisions. Namely, genetic relatedness or kinship plays a key role in the evolution of cooperation in nature \cite{Hamilton1964, Queller1992, West2007a, Clutton-Brock2009a}. Kin selection is pervasive \cite{Dugatkin1997,  Bourke2011}, despite controversies over its role in particular phenomena \cite{Nowak2010a, Nowak2017, Abbot2011, Birch2017, Birch2019, VanVeelen2018}. Indeed, these disagreements indicate the need to investigate the role played by genetic relatedness in each specific scenario \cite{Birch2017}.
Therefore, to address this question, we take such mechanisms into account in the evolutionary dynamics of heuristics selection. 
Namely, we have extended the previous analysis and considered that agents could evaluate an additional variable that accounts whom they are interacting with, specifically, genetic proximity, which is one main mechanism ensuring interactions occur among related individuals \cite{Grafen1990}.

We added to the agents' chromosome a gene $\mathbb{K}$ to account for genetic relatedness with the interacting agent. Operationally, we consider that this kinship relation is given by the Jaccard index of pairs of agents' chromosomes. Note that we are not specifying a method for kin selection, but allowing the heuristics to take into consideration agents similarity when deciding to cooperate or not. Enabling, thus, an estimation of the relevance of genetic relatedness by evaluating the weight organically given to the heuristics' new gene.

\begin{figure}
	\includegraphics[width=0.5\textwidth]{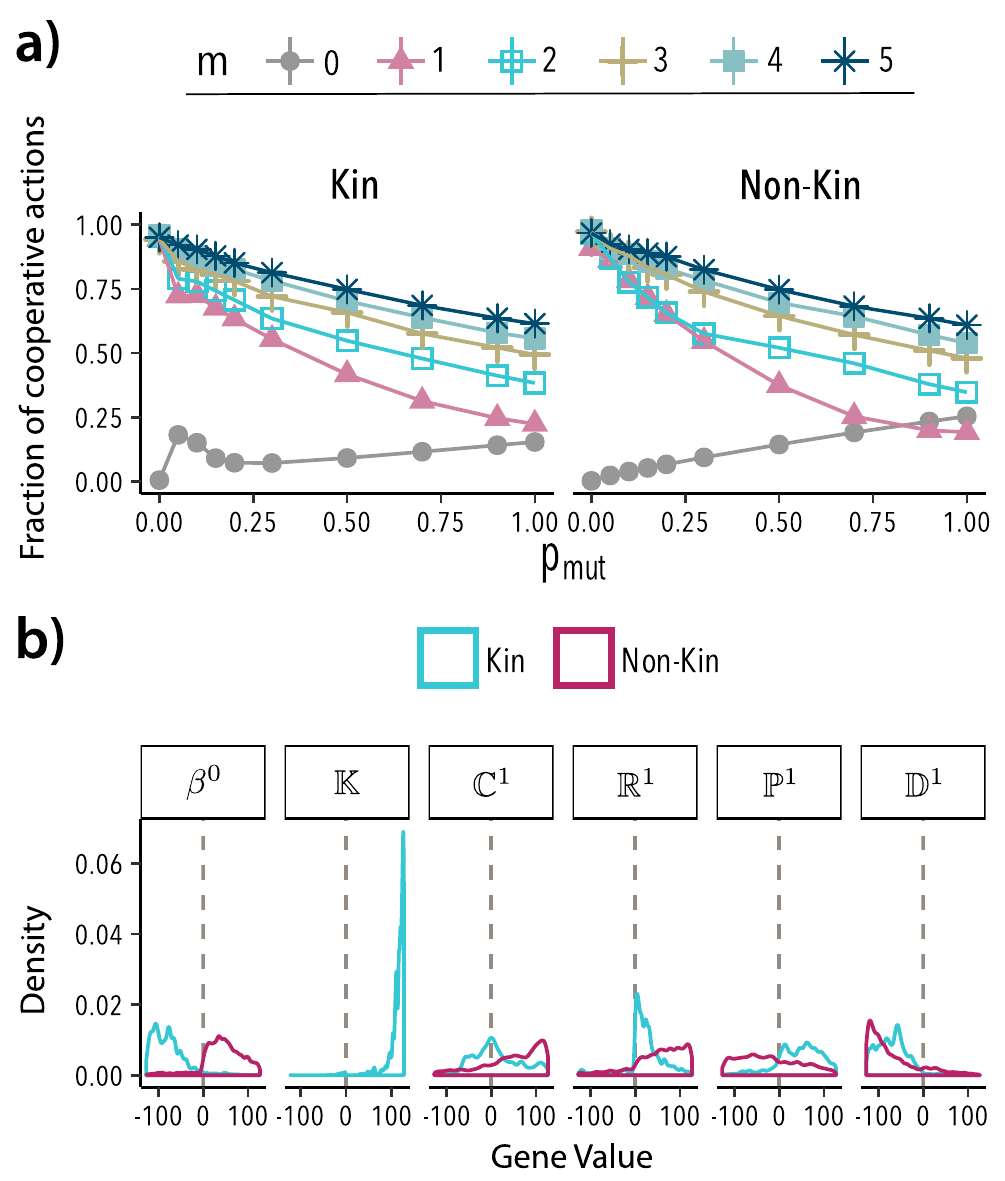}
	\caption{\textbf{\textbf{Evolution of heuristics with kin identification.}} \textbf{A} Fraction of cooperative actions at the steady state as a function of the mutation probability. Colours and shapes correspond for different memory ($m$) values. Averages plus .95 confidence interval of 100 realizations are presented for each mutation ($p_{mut}$) value. \textbf{B} Densities of genes values for simulations on LTT  graphs for $p_{mut}=0.05$ and $m=1$. The vertical dashed lines separate the regions wherein the probability to cooperate would be smaller (gene value smaller than 0) and greater (gene value smaller than 0) than 0.5.}  
	\label{fig:kin}
\end{figure}

Results of simulations on a lattice are presented in Figure \ref{fig:kin}. Figure \ref{fig:kin}A shows the fraction of cooperation at the steady-state both for our previous model (\textit{Non-Kin}) and for the extended model (\textit{Kin}). The evolution leads to similar scenarios in both cases, indicating that the presence of the ($\mathbb{K}$) gene did not enhance nor undermine cooperation significantly, though there is one modest exception. For heuristics without memory ($m=0$) and low mutation, there is a modest increase in the level of cooperation.

Despite the negligible differences in outcomes, there is a substantial effect on agents' chromosomes. Fig. \ref{fig:kin}B shows that including the possibility to weigh gene similarity changes the values of all other genes significantly. For $m=1$, cooperation is strongly determined by the ($\mathbb{K}$) gene, and genes for direct reciprocity and constant response becomes negative or neutral. The latter implies that most agents will not cooperate in one-shot interactions with unrelated individuals, as shown in Supplementary Fig. \ref{fig:SIoneshotKin}, demonstrating a significant difference from the agents without the $\mathbb{K}$ gene.
There still is a mostly positive response for indirect reciprocity and a negative for punishment,  while the weight given to participants payoff inverts. This result points to a compelling message: when heuristics can evaluate genetic relatedness, the ones that do that will have a higher reproduction, therefore resulting in more adapted heuristics.  Nonetheless, information from past interactions is still required, with punishment and reciprocity playing a role.

\section{Conclusions}

Natural selection has shaped the evolution of all sort of life forms. Advantageous strategies endure while others dwindle in a never-ending process of adaptation. Fundamental questions regarding the emergence of cooperative behaviour in social dilemmas have to be studied in the light of evolutionary mechanisms. Undoubtedly, emerging behaviour is intrinsically dependant on the individuals under study, e.g., humans commonly cooperate in large societies composed of unrelated individuals, while groups of animals are hardly greater than a few hundred \cite{Moffett2013}. In particular, variance in humans is especially relevant, as behaviour is deeply affected by the specifics of the interactions and the culture of the individuals \cite{Camerer2003, Henrich2005}. Moreover, given that it is an emergent phenomenon, behaviour can be deeply affected by the complex topology of interactions \cite{Szabo2007b}.  In an attempt to provide a framework for such scenarios, here we explore a model that allows unravelling what could be the drivers of cooperation by a heuristics selection process.

By exploring heuristics that make use of agents behavioural information to stochastically determine their decisions in iterated prisoners' dilemma games across generations, we have shown that, in a feasible environment, evolution will drive heuristics towards cooperation even when defection is expected for pure strategies. In these scenarios, reciprocity and punishment are the main ingredients of cooperators' decision-making, and most strategies will follow conditional cooperation.  The fraction of cooperative decisions decreases with an increase in the mutation rate, nonetheless, for small mutation rates the system reaches a cooperative equilibrium. Without mutation, the configuration of the initial state is critical and the system can get trapped in equilibria of meagre cooperation. Increasing the memory of individuals also increases the fraction of cooperation, suggesting that heuristics with more resources are more cooperative.  These aggregate results are indistinguishable from a version of the model wherein agents have, in addition to behavioural information, access to their similarity with others (which mimics genetic relatedness). For this latter scenario, the level of cooperation at the macroscopic level remains roughly the same. Important enough, however, at the level of individuals, chromosomes change significantly and cooperation is given through a kin identification process. 

Therefore, when agents discriminate their kin, reciprocity loses much of its importance, which is especially insightful given the behaviour observed in nature. Kin selection is arguably the most important mechanism behind cooperation in non-human animals, while reciprocity is uncommon \cite{West2007a, Clutton-Brock2009a}. Our result suggests that in order for reciprocity to be dominant, perfect kin discrimination cannot exist, which suggest that figuring out the interplay between both mechanisms is crucial for understanding human evolution. Moreover, agents evolved in each condition presented a different expected response in one-shot games with unrelated individuals: cooperation is likely without the kin discrimination gene, while the majority of agent will defect when they can discriminate their genetic similarity. 

To round off, we note that heuristics will adapt according to the information that they have access to, and they can change significantly according to the variables available. Surprisingly, despite changes in methods, cooperation is more likely than exploitation, due to reciprocity \cite{Trivers1971, Alexander1987} or to kin selection \cite{Hamilton1964}. This suggests that even if individuals have limited cognitive capacities (a small memory weighed by a rather inexpensive function),  cooperative heuristics can have higher reproduction rates and be pervasive. 
However, extrapolations have to be made with caution. As it is often the case of works in evolutionary game theory, our model sidesteps important details from biology and cognitive sciences \cite{VanCleve2020}.  Future work should explore the intersection between moral and material values and how it influences heuristics \cite{Bowles2011}, and how selection works in more complex scenarios, for instance, when higher cognition has higher associated costs \cite{Seoane2018}. Moreover, our approach could be used to understand how cultural characteristics \cite{Henrich2005, Boyd2018} drive cooperation in different directions by modelling proper environmental variables, and whether costly punishment could sustain large scale cooperation \cite{Fehr2002a}. We plan to explore this and similar questions next.

\begin{acknowledgments} 
We acknowledge partial support from Project No. UZ-I-2015/022/PIP, the Government of Arag\'on, Spain, FEDER Funds, through Grant No. E36-17R to FENOL, and from MINECO and FEDER funds (Grant No. FIS2017-87519-P). YM acknowledges support from Intesa Sanpaolo Innovation Center. The funders had no role in study design, data collection, and analysis, decision to publish, or preparation of the manuscript.
\end{acknowledgments}

%\bibliographystyle{vancouver}
%\bibliography{references} % TODO - prepare proper file before sub 

\clearpage

\FloatBarrier
\newpage

\begin{center}
\textbf{\large Supplemental Materials: Dynamics of heuristics selection for cooperative behavior}
\end{center}
%%%%%%%%%% Merge with supplemental materials %%%%%%%%%%
%%%%%%%%%% Prefix a "S" to all equations, figures, tables and reset the counter %%%%%%%%%%
\setcounter{section}{0}
\setcounter{equation}{0}
\setcounter{figure}{0}
\setcounter{table}{0}
\setcounter{page}{1}
\makeatletter
\renewcommand{\theequation}{S\arabic{equation}}
\renewcommand{\thefigure}{S\arabic{figure}}
\renewcommand{\thetable}{S\arabic{table}}
% \renewcommand{\bibnumfmt}[1]{[S#1]}
% \renewcommand{\citenumfont}[1]{S#1}
%%%%%%%%%% Prefix a "S" to all equations, figures, tables and reset the counter %%%%%%%%%%

\section{Other payoff values} \label{sec:payoffT}

\begin{figure}
	\includegraphics[width=0.85\textwidth]{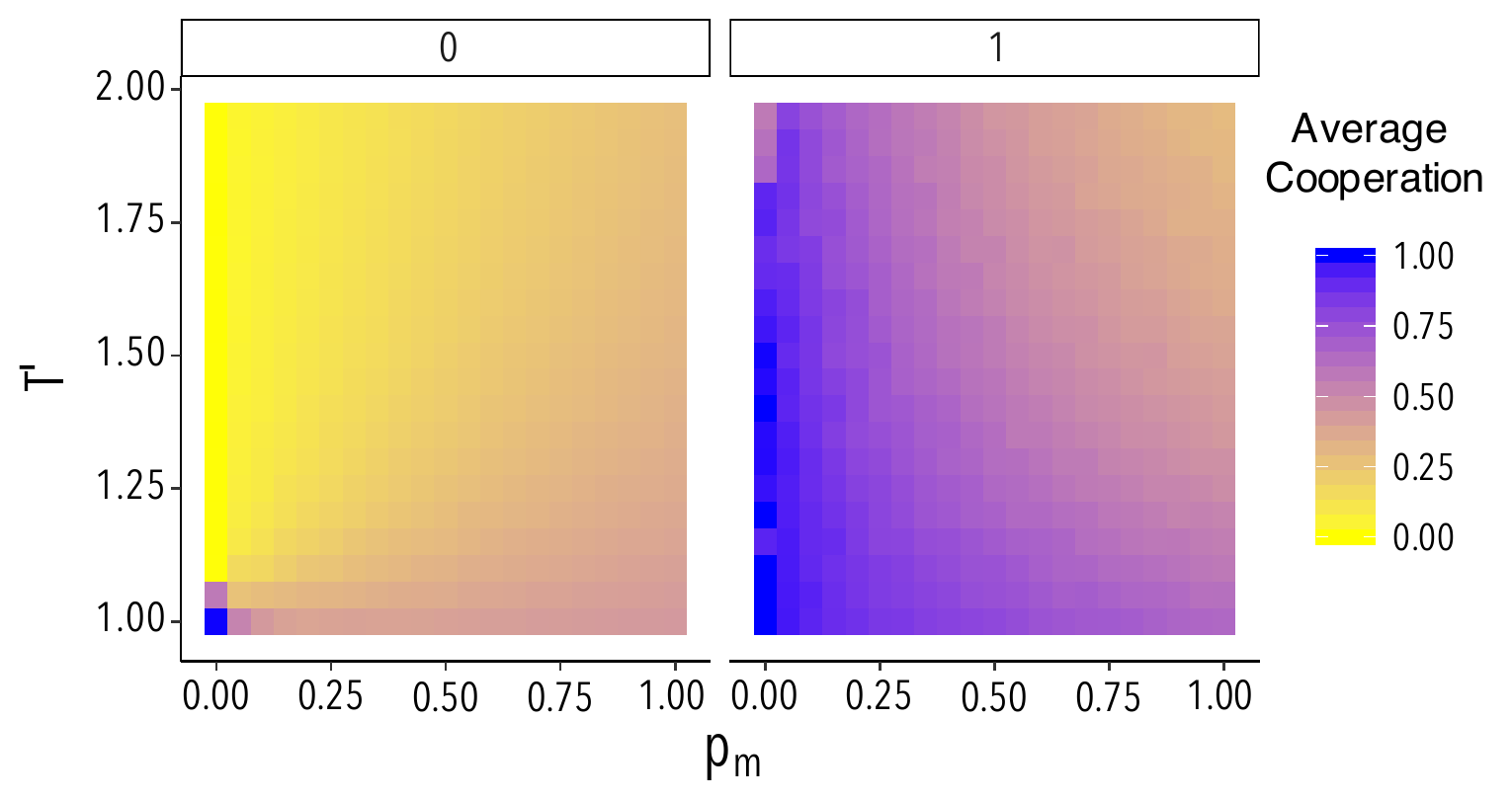}
	\caption{\textbf{Average cooperation at the stationary state}. The left panel shows results for $m=0$ and the right panel for $m=1$. 100 simulations were done for each $p_m$ and $T'$ combination. Colour coding corresponds to the average over all realizations and varies from blue (1) to yellow (0).  }  \label{fig:SIheatmap}
\end{figure}

\begin{figure}
	\includegraphics[width=0.5\textwidth]{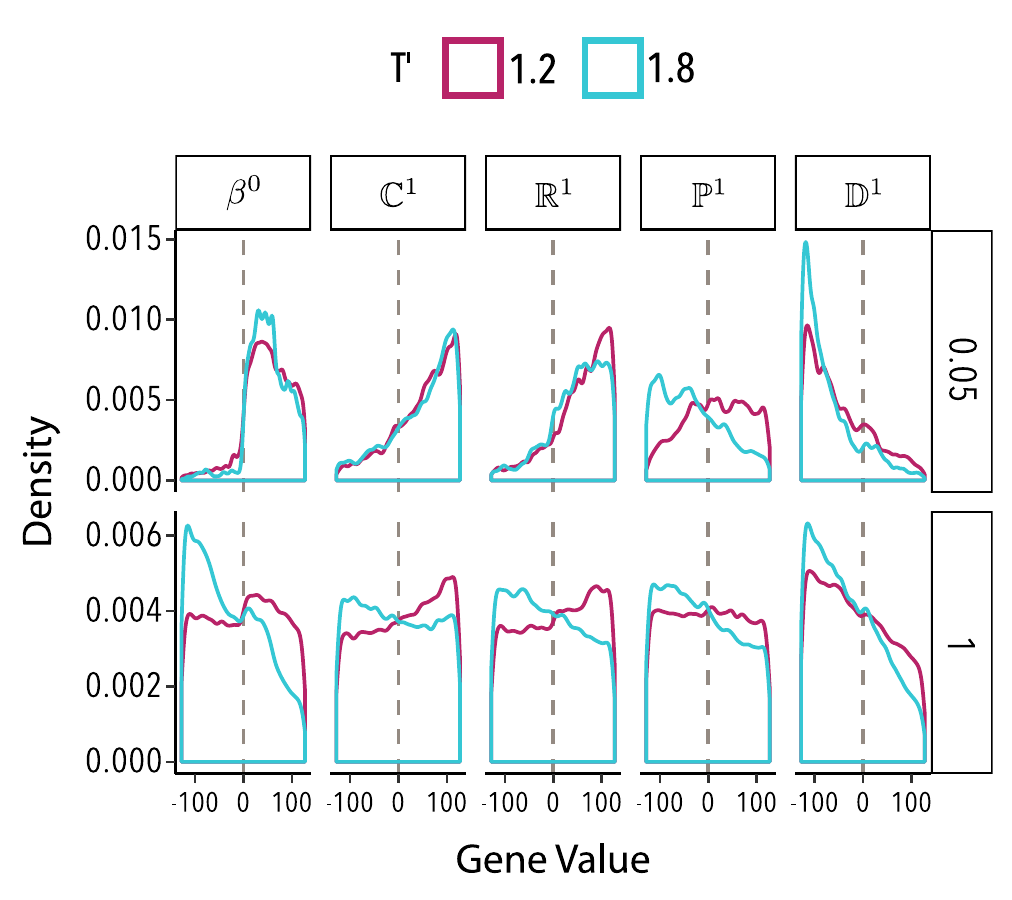}
	\caption{\textbf{\textbf{Distribution of genes' expressed values.}} Densities of genes values for simulations on a lattice for $m=1$. Top panels show distributions for $p_{mut}=0.05$ and bottom panels for $p_{mut}=1$. The vertical dashed lines separate regions wherein the marginal probability to cooperate would be smaller (negative gene values) and greater (positive gene value) than 0.5.}  \label{fig:STgeneDist}
\end{figure}

To ensure that our results are robust with respect to differences in the payoff values, we ran simulations for different values of the temptation parameter $T$. To make our results comparable to previous work, we used the one-dimensional parametrization of payoffs used by Nowak et. al \cite{Nowak1992}. In this version, $R=1$, $P=\epsilon$, $S=0$, and $T$ varies from 1 to 2, with $\epsilon$ being a value close to zero. As we consider normalized versions, the payoff here is defined by $T=T'/\braket{k}; R=1/\braket{k}; P=0.01/\braket{k}; S = 0$, with $T'$ varying from 1 to 2. Results for memory 0 and 1 are shown in Fig. \ref{fig:SIheatmap}. The results show that without memory, cooperation is only attainable when $T'=1$ and low mutation. However, when agents have memory of their last interaction, cooperation endures even when the temptation to defect is around 2. 
As an illustration, the distributions of gene values for $T'$ values of 1.2 and 1.8 are shown in Fig \ref{fig:STgeneDist}. They follow a close pattern to the ones shown in Fig. \ref{fig:geneDist} in the main text, indicating the equivalence of both results. This shows that our results are robust across a broad range of parameter values. 

\section{Heuristics Classification} \label{sec:classification}

Heuristics are classified according to two basic strategies: \textit{Pure Cooperation} and \textit{Pure Defection}. These two strategies always cooperate and always defect, respectively.  Table \ref{tb:history} illustrates the variables contained in the memory of agent $u$ with respect to a player $v$, corresponding to the two pure strategies for $m=1$. All the values are given straightforwardly, expect for  $\pi_v^1$. Payoffs values are more complicated, as they depend on the players with whom they are playing with, which we cannot define a priori. We decided to use the average payoff of individuals which cooperated and defected with all their neighbours for the pure cooperator and pure defector, respectively. Therefore, $\braket{\pi_{C}} = \braket{\pi_{i}^t} \forall i \in R_1, \forall t \in [1,100]$ and 
$\braket{\pi_{D}} = \braket{\pi_{i}^t} \forall i \in R_0, \forall t \in [1,100]$, wherein $R_1$ (resp. $R_0$) corresponds to the set of agents which cooperated with all (resp. none) of their neighbours in the last time step.

The activation function (Eq. \ref{eq:decisionf} in the main text) of an agent results in the probability to cooperate with the Pure Cooperator ($\rho_C$) and the Pure Defector ($\rho_D$). We then, use the threshold $\sigma$ to divide the plane $(\rho_C, \rho_D)$. Namely, we designate as cooperation when $\rho > (1 - \sigma)$, defection as $\rho < \sigma$, and random when $\sigma \leq \rho \leq (1-\sigma)$.  This process results in the set of strategies given in Table \ref{tb:strategies} of the main text. Therefore, a precise version of this table would correspond to the one shown in Table \ref{tb:strategiesSI}.

\begin{table}
	\centering
	\begin{tabular}{|c|c|c|}
		\hline 
		& \textbf{Pure Cooperator} & \textbf{Pure Defector}   \\ 
		\toprule
		$C_{v,u}^1$ & 1 & 0   \\ \hline     
		$R_{v, u}^1$ & 1 & 0 \\ \hline 
		$\pi_v^1$ & $ \braket{\pi_{C}}$ & $\braket{\pi_{D}}$ \\ \hline     
		$D_{v, u}^1$ &  0 & 1 \\ \hline
	\end{tabular} 
	\caption{\textbf{Pure strategies memory.} Past history  of the two basic strategies according to what would have been played by them for $m=1$.}
	\label{tb:history}
\end{table}

\begin{table}
	\centering-
	\begin{tabular}{|c|c|c|}
		\hline 
		& \textbf{Pure Cooperator} & \textbf{Pure Defector}   \\ 
		\toprule
		FC & $\rho > (1 - \sigma)$  & $\rho > (1 - \sigma)$  \\ \hline 
		FD & $\rho < \sigma$ & $\rho < \sigma$ \\ \hline  
		CC & $\rho > (1 - \sigma)$  & $\rho < \sigma$   \\ \hline 
		GCC & $\rho > (1 - \sigma)$  & $\sigma \leq \rho \leq (1-\sigma)$  \\          \hline
		CD & $\sigma \leq \rho \leq (1-\sigma)$  & $\rho < \sigma$ \\  \hline
		Bully & $\rho < \sigma$ & $\rho > (1 - \sigma)$ \\  \hline
		Random &  $\sigma \leq \rho \leq (1-\sigma)$  & $\sigma \leq \rho \leq (1-\sigma)$ \\   \hline
		%        Undefined & ? & ? \\  \hline
	\end{tabular} 
	\caption{\textbf{Classification of heuristics according to their responses to the two pure strategies}: pure defector and pure cooperator. $\rho$ corresponds to the resulting probability when playing with the corresponding pure strategy.}
	\label{tb:strategiesSI}
\end{table}

\begin{figure}
	\includegraphics[width=0.48\textwidth]{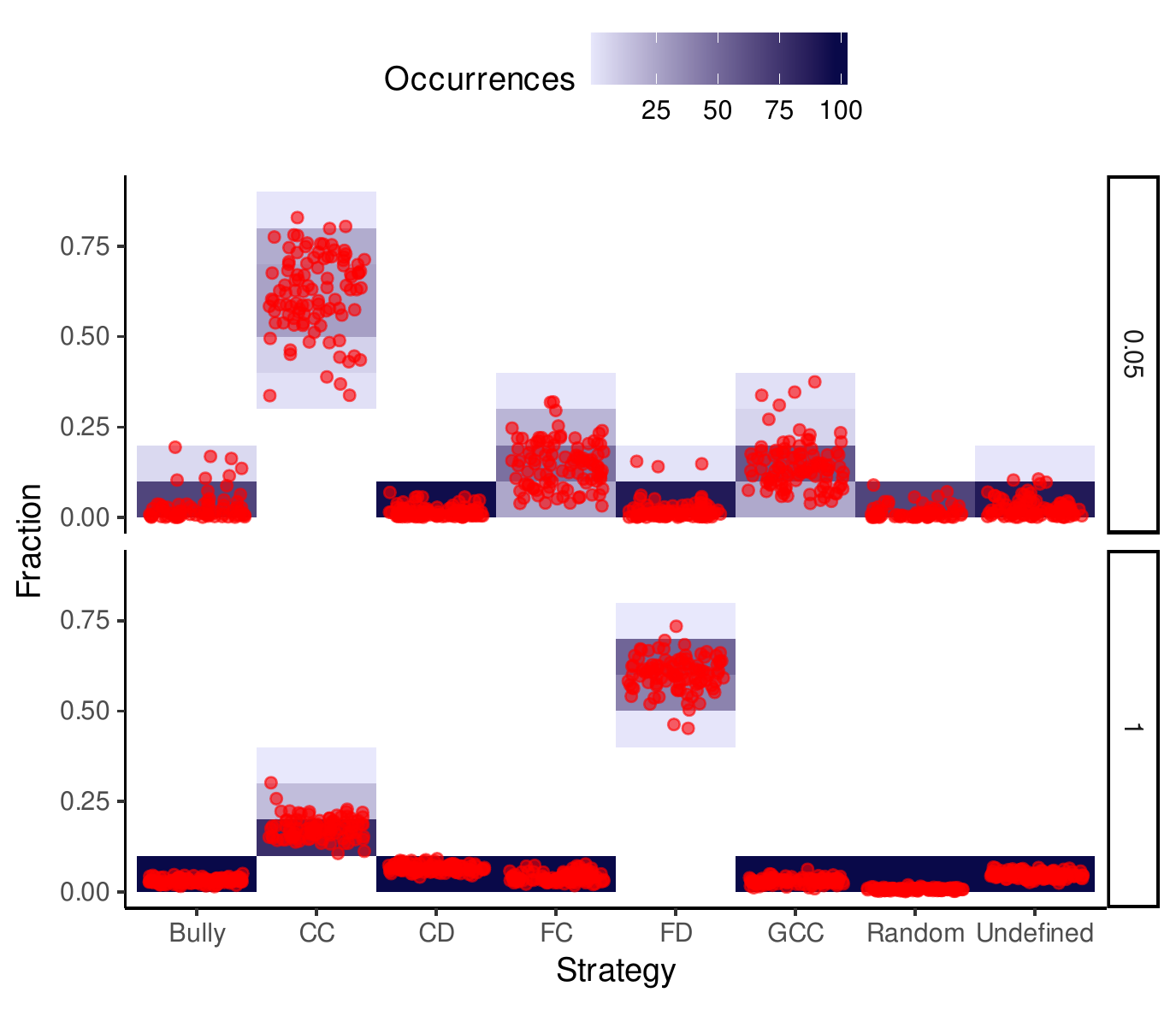}
	\caption{\textbf{\textbf{Emerging Strategies in RRN}} Frequency of each strategy in realizations for RRN networks and $p_{mut}=0.05$, $\sigma=0.3$ and $m=1$. Each red dot corresponds to the fraction of the strategy in a simulation and the histogram of fractions for each strategy is shown vertically, with darkest colours representing a higher number of occurrences. }  \label{fig:SIstrategies}
\end{figure}

\section{Extended model in Random Regular Networks}

In this section, we present the results of the extended model with the Kinship parameter ran on RRN graphs. At variance with the model in a lattice, when there is no mutation,  the fraction of cooperative actions can be different from zero, as it is also shown in the time evolution figures. This demonstrates how important the kin identification mechanism can be in an adequate environment. With mutation, the macroscopic results are equivalent to the results in a lattice and in an RRN without the extension. Furthermore, when agents have had access to memory the results are equivalent to the ones obtained in a lattice, including the distribution of gene values for $m=1$. 

\begin{figure}
	\includegraphics[width=0.48\textwidth]{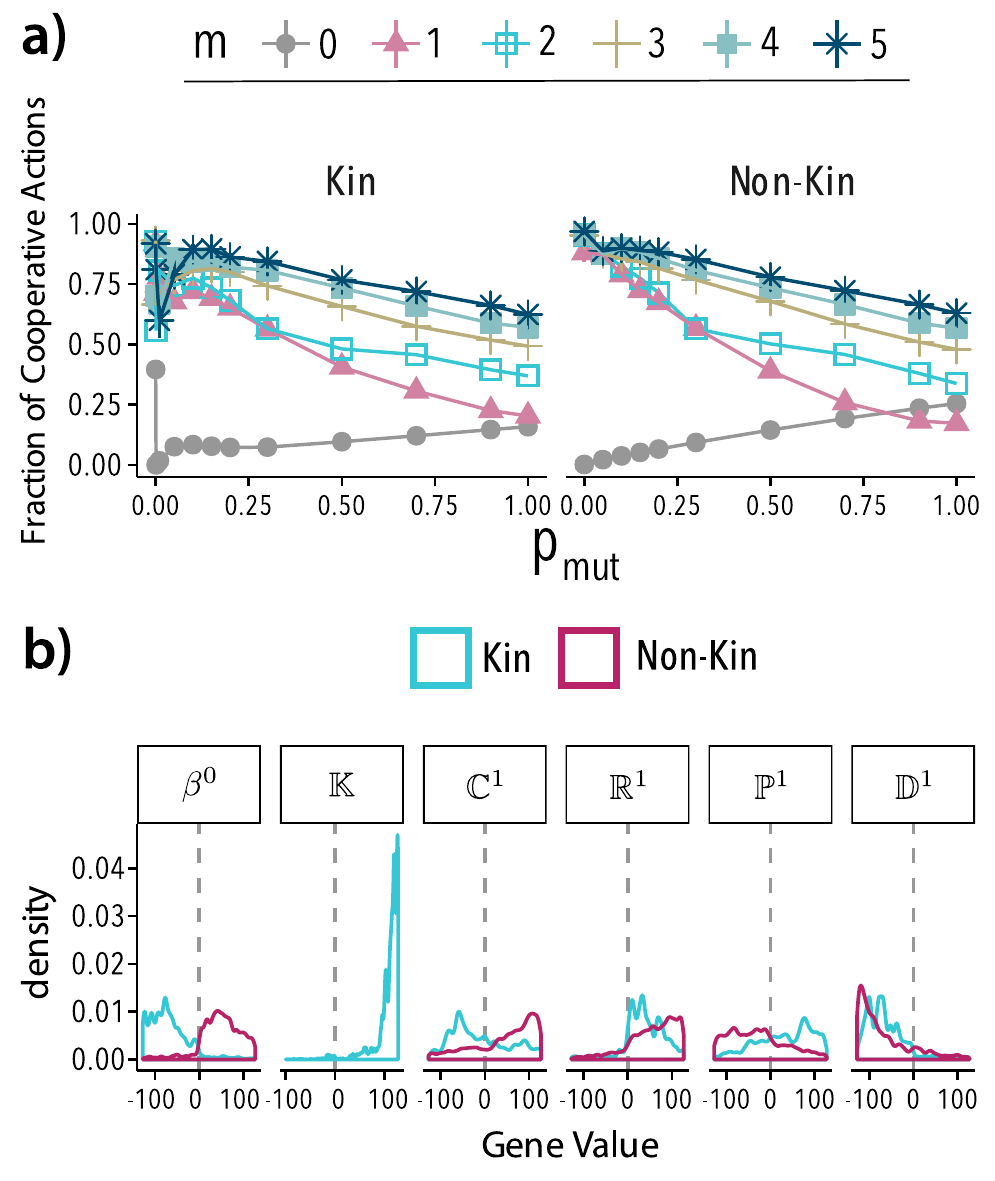}
	\caption{\textbf{\textbf{Evolution of heuristics with kin identification on RRN .}} \textbf{A} Fraction of cooperative actions at the steady state as a function of the mutation probability. Colours and shapes correspond for different memory ($m$) values. Averages plus .95 confidence interval of 100 realizations are presented for each mutation ($p_{mut}$) value. \textbf{B} Densities of genes values for simulations on RRN  graphs for $p_{mut}=0.05$ and $m=1$. The vertical dashed lines separate the regions wherein the probability to cooperate would be smaller (gene value smaller than 0) and greater (gene value smaller than 0) than 0.5. 
	}  \label{fig:RRNkin}
\end{figure}

\section{Time Evolution} \label{sec:timeEvo}

\begin{figure}
	\includegraphics[width=0.78\textwidth]{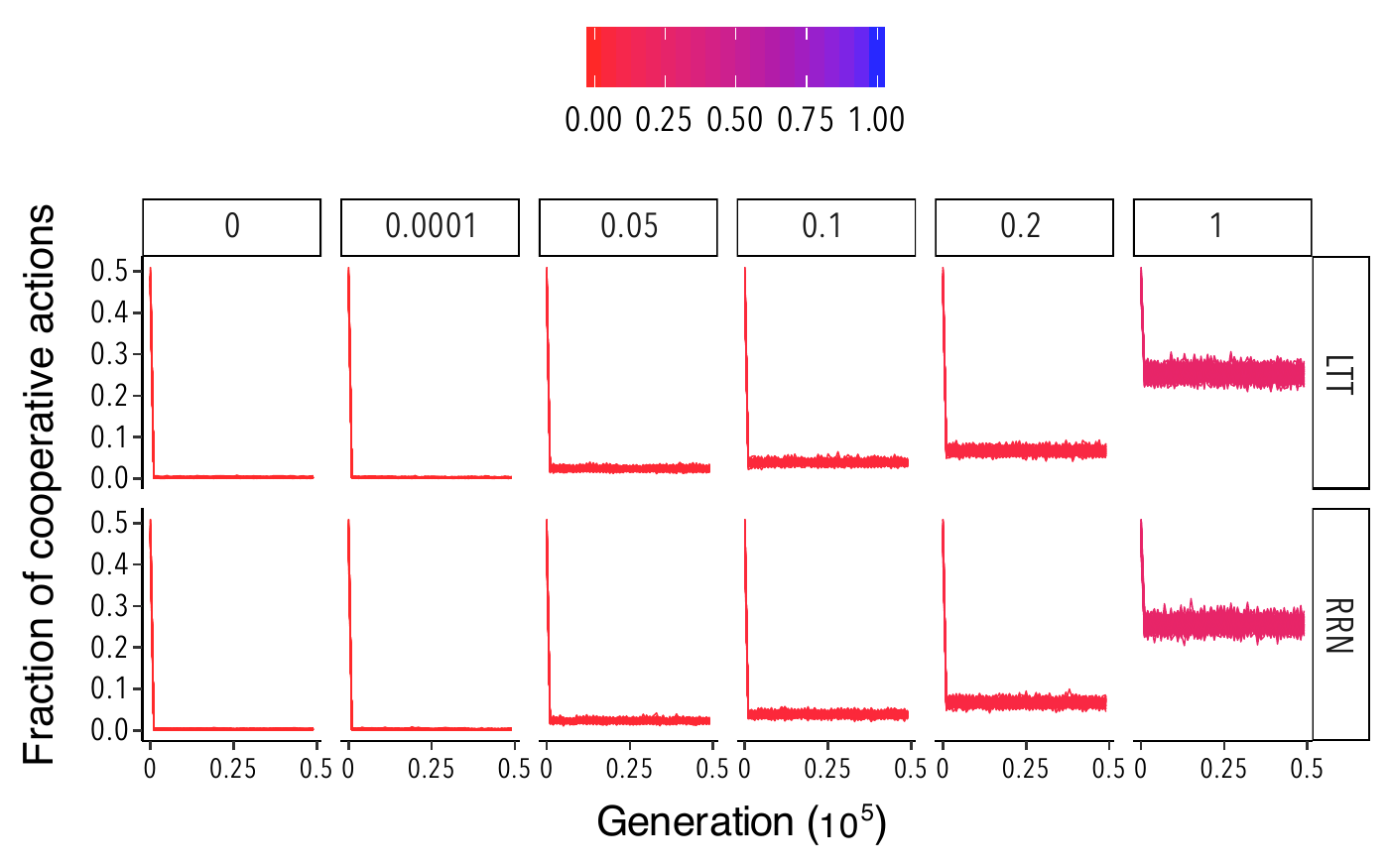}
	\caption{\textbf{Fraction of cooperative actions at the end of each generation}. Columns correspond to different mutation values (0, 0.0001, 0.05, 0.1, 0.2, 1) and horizontal panels to different models (LTT, RRN). Agents have memory $m=0$ and 100 realizations were performed for each mutation value and network. The colours of the lines correspond to the average of the last 1000 time steps.}  \label{fig:SIperC0}
\end{figure}

\begin{figure}
	\includegraphics[width=0.78\textwidth]{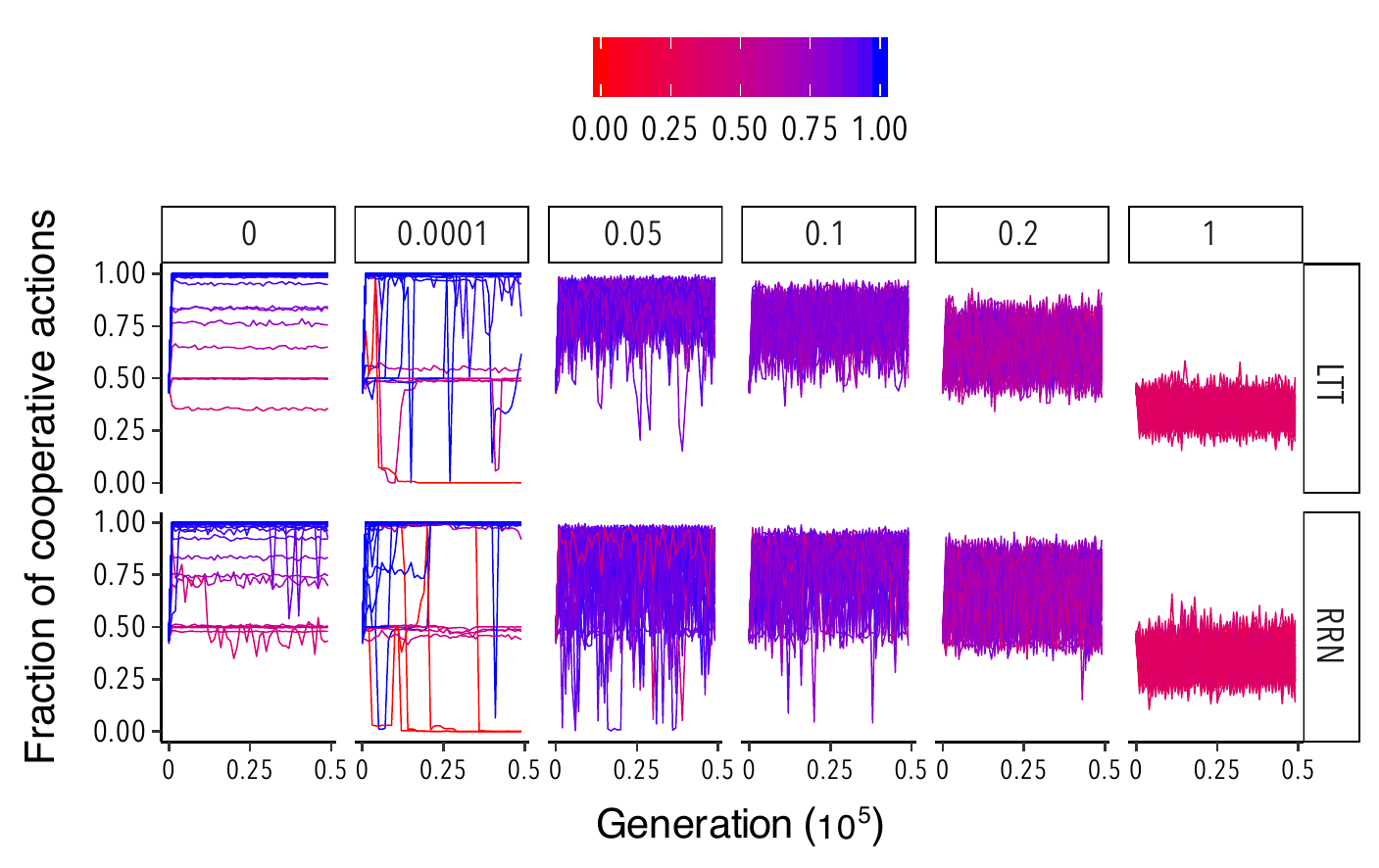}
	\caption{\textbf{Fraction of cooperative actions at the end of each generation}. Columns correspond to different mutation values (0, 0.0001, 0.05, 0.1, 0.2, 1) and horizontal panels to different models (LTT, RRN). Agents have memory $m=2$ and 100 realizations were performed for each mutation value and network. The colours of the lines correspond to the average of the last 1000 time steps.}  \label{fig:SIperC2}
\end{figure}

\begin{figure}
	\includegraphics[width=0.78\textwidth]{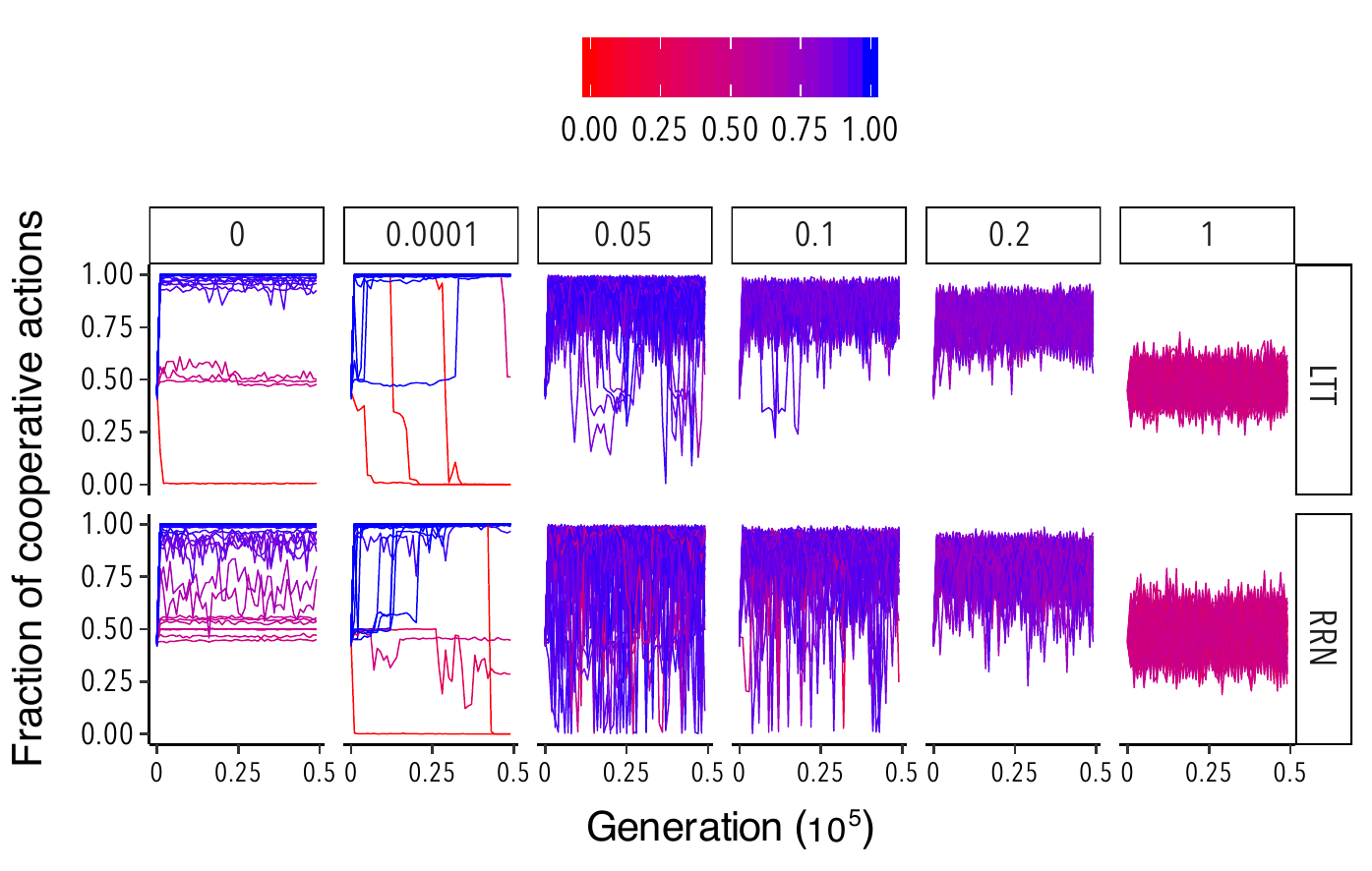}
	\caption{\textbf{Fraction of cooperative actions at the end of each generation}. Columns correspond to different mutation values (0, 0.0001, 0.05, 0.1, 0.2, 1) and horizontal panels to different models (LTT, RRN). Agents have memory $m=3$ and 100 realizations were performed for each mutation value and network. The colours of the lines correspond to the average of the last 1000 time steps.}  \label{fig:SIperC3}
\end{figure}

\begin{figure}
	\includegraphics[width=0.78\textwidth]{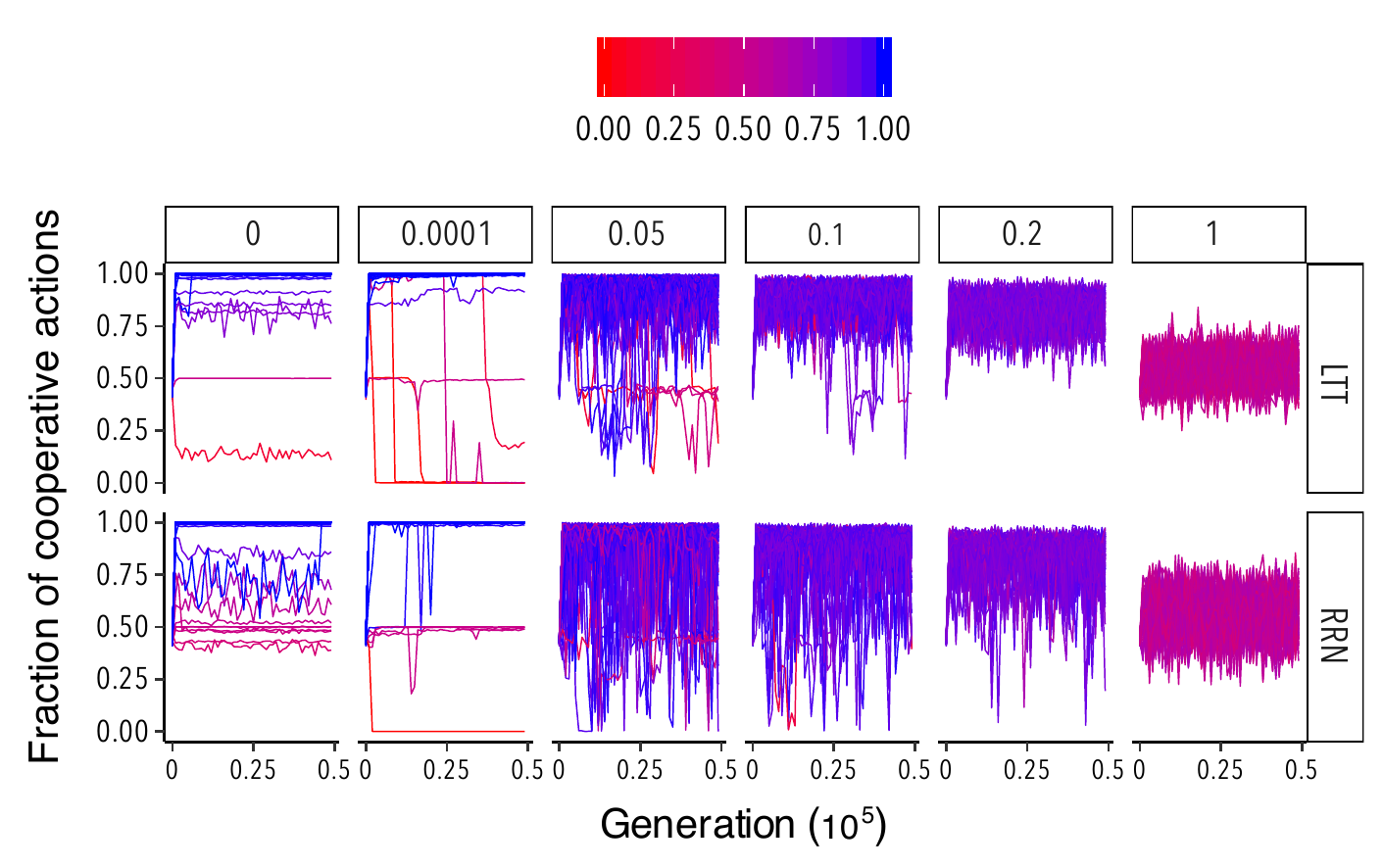}
	\caption{\textbf{Fraction of cooperative actions at the end of each generation}. Columns correspond to different mutation values (0, 0.0001, 0.05, 0.1, 0.2, 1) and horizontal panels to different models (LTT, RRN). Agents have memory $m=4$ and 100 realizations were performed for each mutation value and network. The colours of the lines correspond to the average of the last 1000 time steps.}  \label{fig:SIperC4}
\end{figure}

\begin{figure}
	\includegraphics[width=0.78\textwidth]{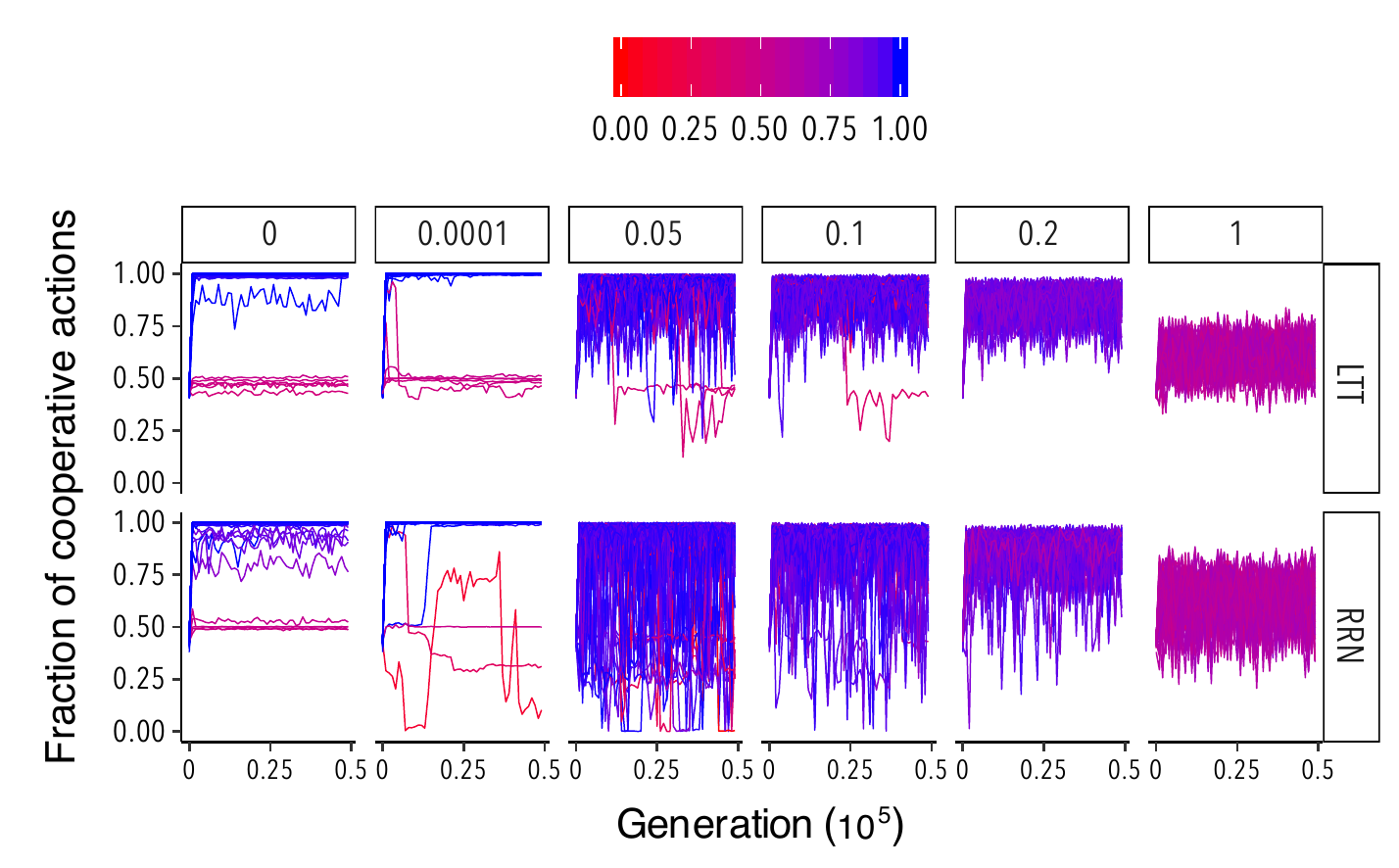}
	\caption{\textbf{Fraction of cooperative actions at the end of each generation}. Columns correspond to different mutation values (0, 0.0001, 0.05, 0.1, 0.2, 1) and horizontal panels to different models (LTT, RRN). Agents have memory $m=5$ and 100 realizations were performed for each mutation value and network. The colours of the lines correspond to the average of the last 1000 time steps.}  \label{fig:SIperC5}
\end{figure}

\begin{figure}
	\includegraphics[width=0.78\textwidth]{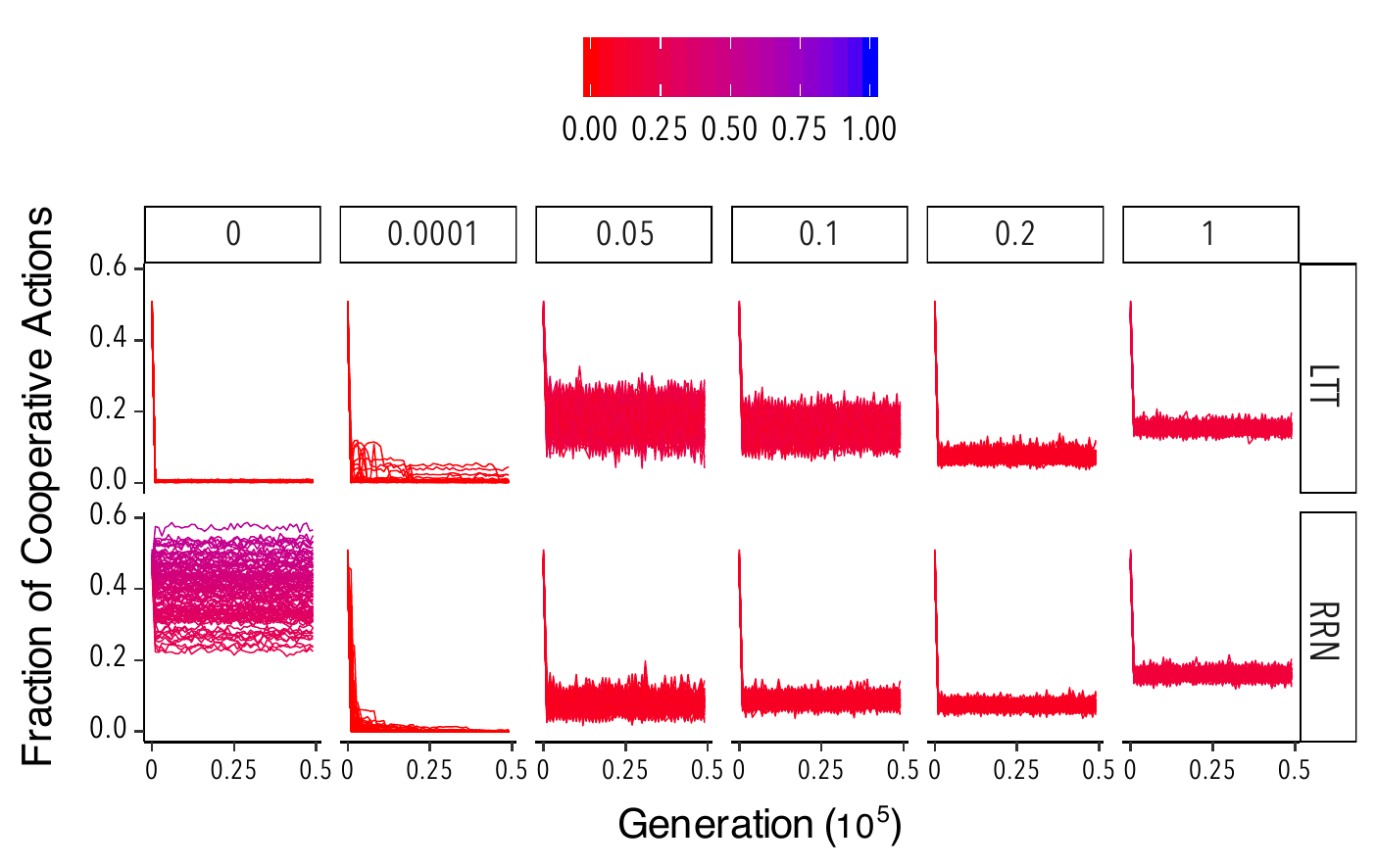}
	\caption{\textbf{Fraction of cooperative actions at the end of each generation for the model with the kin identification gene}. Columns correspond to different mutation values (0, 0.0001, 0.05, 0.1, 0.2, 1) and horizontal panels to different models (LTT, RRN). Agents have memory $m=$ and 100 realizations were performed for each mutation value and network. The colours of the lines correspond to the average of the last 1000 time steps.}  \label{fig:SIperCKin0}
\end{figure}

\begin{figure}
	\includegraphics[width=0.78\textwidth]{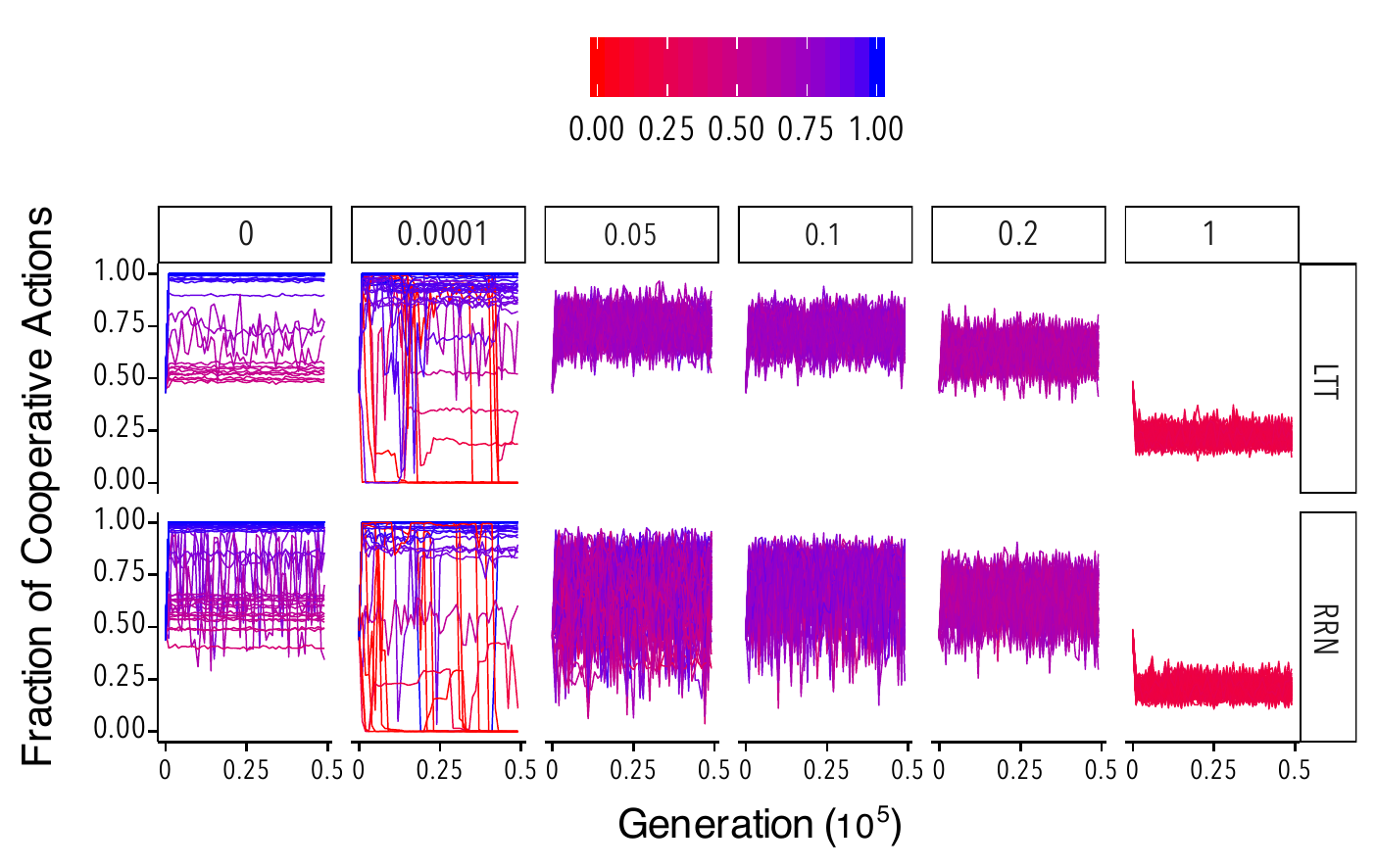}
	\caption{\textbf{Fraction of cooperative actions at the end of each generation for the model with the kin identification gene}. Columns correspond to different mutation values (0, 0.0001, 0.05, 0.1, 0.2, 1) and horizontal panels to different models (LTT, RRN). Agents have memory $m=1$ and 100 realizations were performed for each mutation value and network. The colours of the lines correspond to the average of the last 1000 time steps.}  \label{fig:SIperCKin1}
\end{figure}

\begin{figure}
	\includegraphics[width=0.78\textwidth]{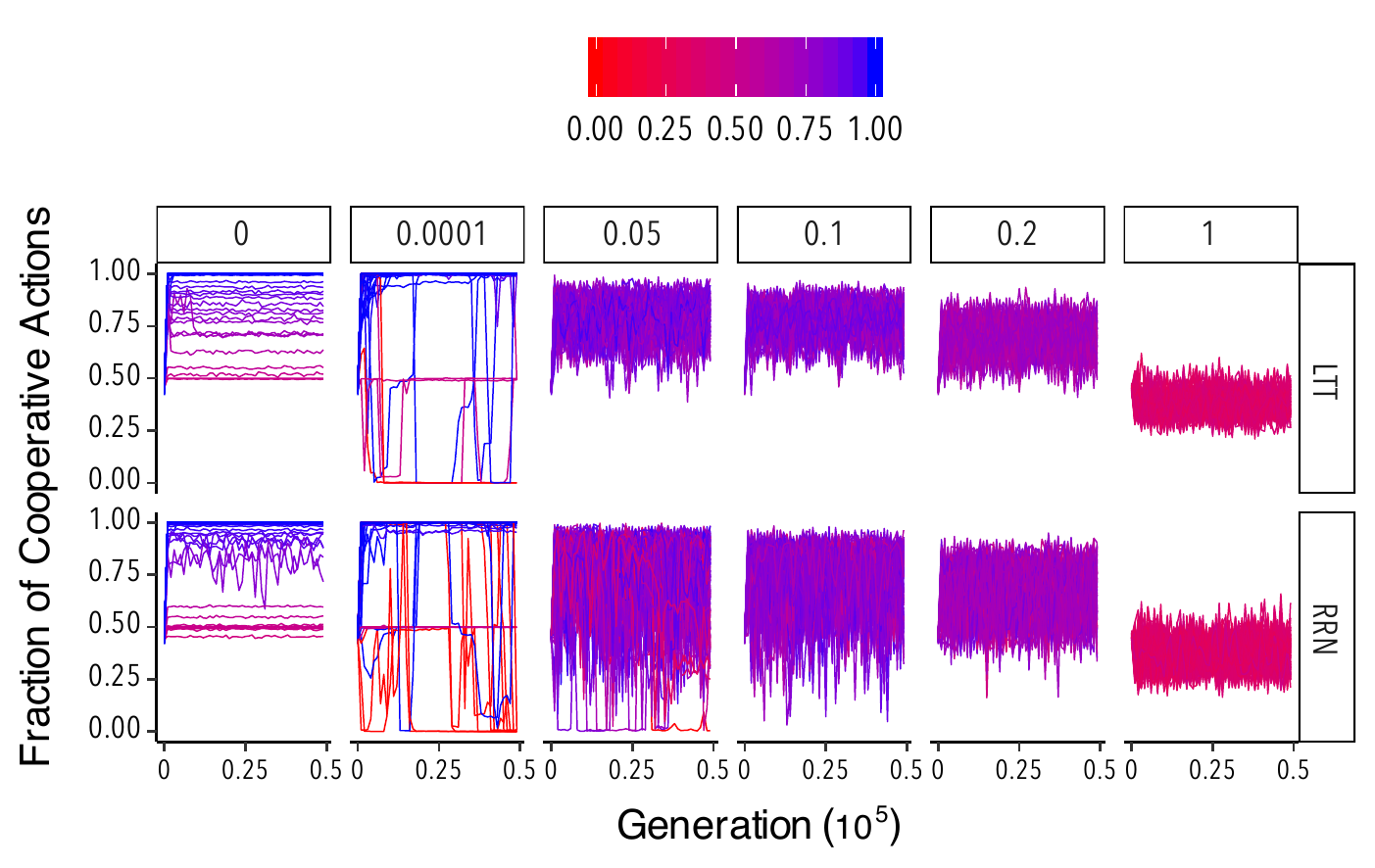}
	\caption{\textbf{Fraction of cooperative actions at the end of each generation for the model with the kin identification gene}. Columns correspond to different mutation values (0, 0.0001, 0.05, 0.1, 0.2, 1) and horizontal panels to different models (LTT, RRN). Agents have memory $m=2$ and 100 realizations were performed for each mutation value and network. The colours of the lines correspond to the average of the last 1000 time steps.}  \label{fig:SIperCKin2}
\end{figure}

\begin{figure}
	\includegraphics[width=0.78\textwidth]{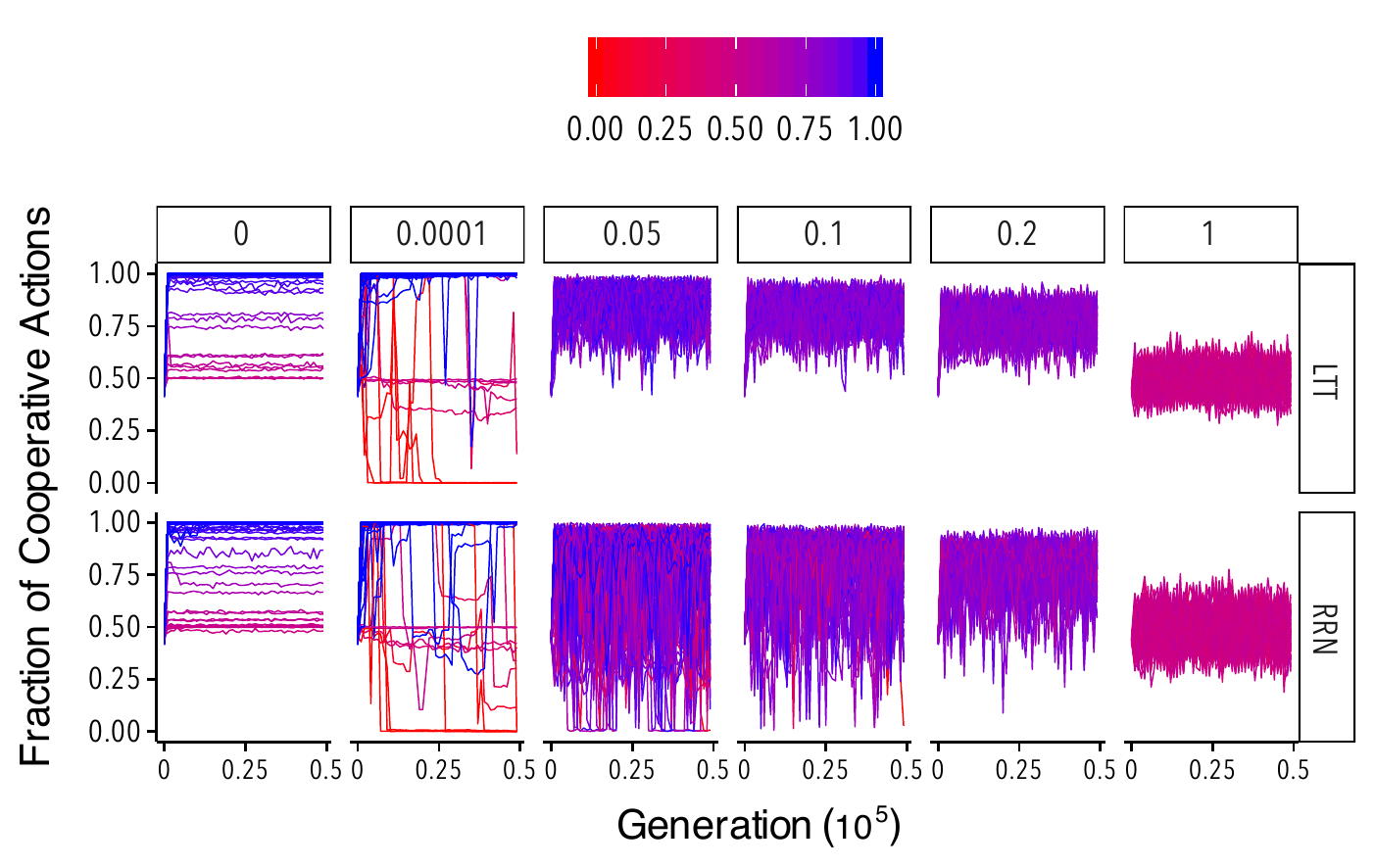}
	\caption{\textbf{Fraction of cooperative actions at the end of each generation for the model with the kin identification gene}. Columns correspond to different mutation values (0, 0.0001, 0.05, 0.1, 0.2, 1) and horizontal panels to different models (LTT, RRN). Agents have memory $m=3$ and 100 realizations were performed for each mutation value and network. The colours of the lines correspond to the average of the last 1000 time steps.}  \label{fig:SIperCKin3}
\end{figure}

\begin{figure}
	\includegraphics[width=0.78\textwidth]{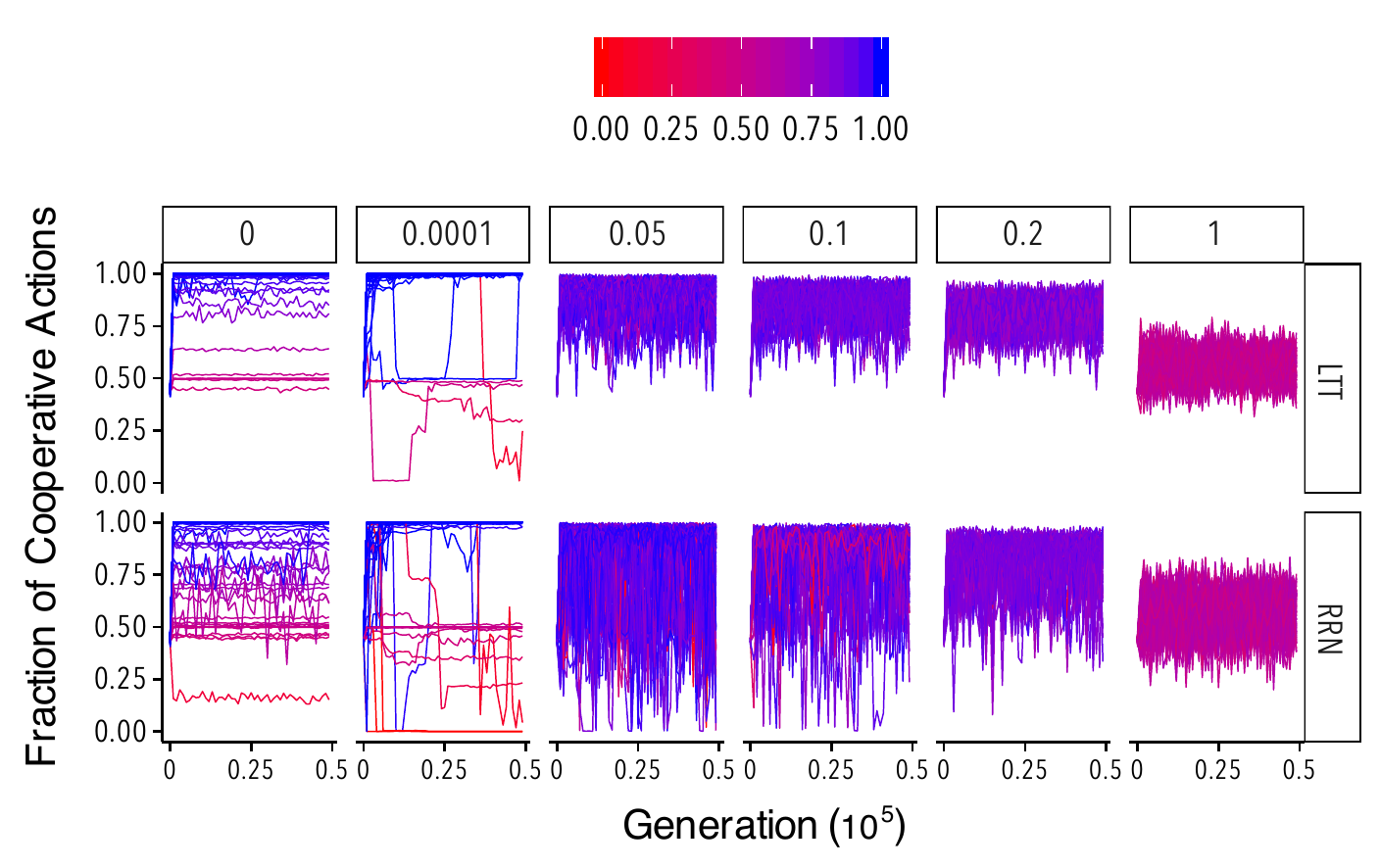}
	\caption{\textbf{Fraction of cooperative actions at the end of each generation for the model with the kin identification gene}. Columns correspond to different mutation values (0, 0.0001, 0.05, 0.1, 0.2, 1) and horizontal panels to different models (LTT, RRN). Agents have memory $m=4$ and 100 realizations were performed for each mutation value and network. The colours of the lines correspond to the average of the last 1000 time steps.}  \label{fig:SIperCKin4}
\end{figure}

\begin{figure}
	\includegraphics[width=0.78\textwidth]{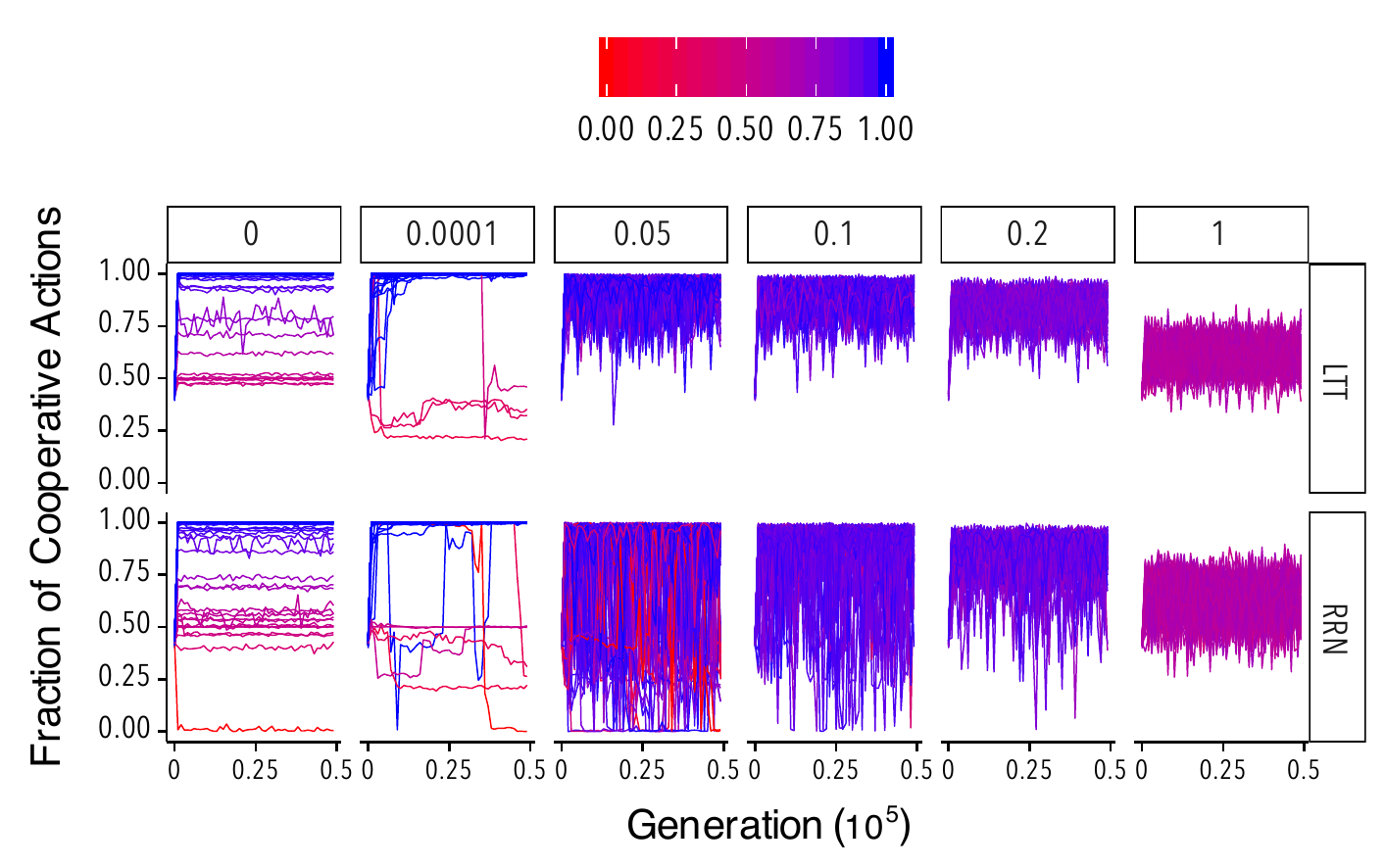}
	\caption{\textbf{Fraction of cooperative actions at the end of each generation for the model with the kin identification gene}. Columns correspond to different mutation values (0, 0.0001, 0.05, 0.1, 0.2, 1) and horizontal panels to different models (LTT, RRN). Agents have memory $m=5$ and 100 realizations were performed for each mutation value and network. The colours of the lines correspond to the average of the last 1000 time steps.}  \label{fig:SIperCKin5}
\end{figure}

Time evolution curves of the main model executions for memory values of 0,2,3,4, and 5 are shown in Figures \ref{fig:SIperC0}, \ref{fig:SIperC2}, \ref{fig:SIperC3}, \ref{fig:SIperC4}, and \ref{fig:SIperC5}, respectively. Figures \ref{fig:SIperCKin0}, \ref{fig:SIperCKin1}, \ref{fig:SIperCKin2}, \ref{fig:SIperCKin3}, \ref{fig:SIperCKin4}, and \ref{fig:SIperCKin5} show the
time evolution curves of realizations of the model with the addition of the kin identification gene for memory values of 0,1,2,3,4, and 5, respectively.


% Time evolution of the Kin Model

\section{One-shot responses} \label{sec:oneshot}

\begin{figure}
	\includegraphics[width=0.5\textwidth]{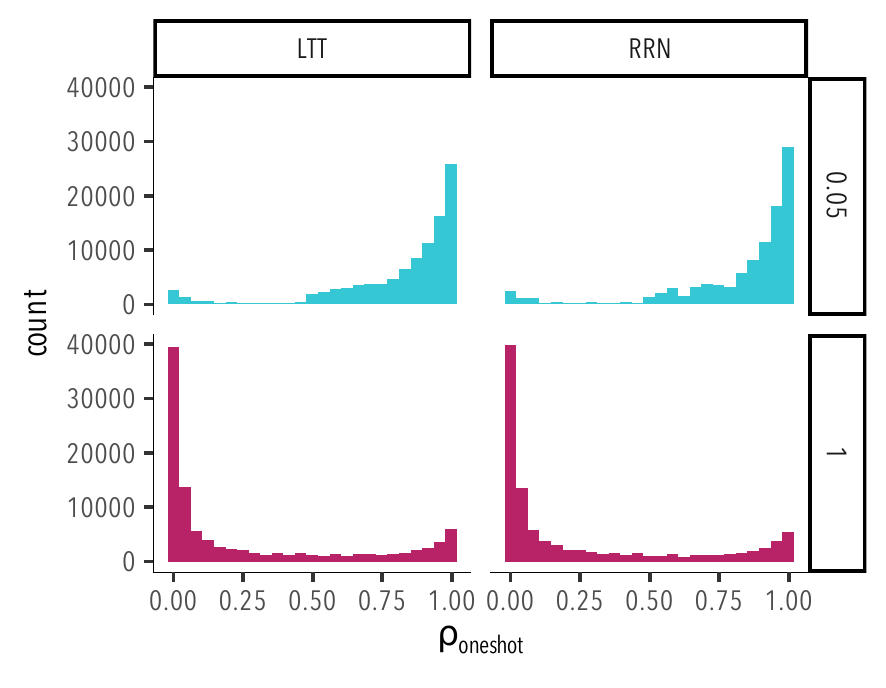}
	\caption{\textbf{Cooperation probability in one-shot games}. Distributions are calculated over agents of the final generation for $m=1$ in a lattice (left panel) and in random regular networks (right panel). Probability is calculated considering that agents do not have access to other participants information, thus, only $\beta^0$ is used in the sigmoid. Top panels show distributions for $p_{mut}=0.05$ and bottom panels for $p_{mut}=1$. }  \label{fig:SIoneshot}
\end{figure}

\begin{figure}
	\includegraphics[width=0.5\textwidth]{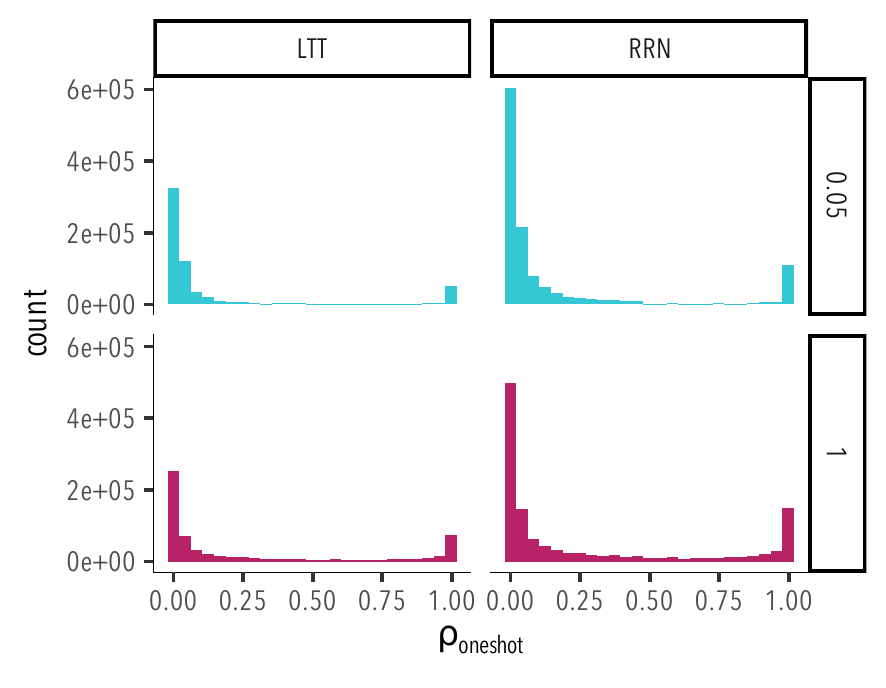}
	\caption{\textbf{Cooperation probability in one-shot games for the model with the kin identification gene}. Distributions are calculated over agents of the final generation for $m=1$ in a lattice (left panel) and in random regular networks (right panel). Probability is calculated considering that agents do not have access to other participants information, thus, only $\beta^0$ is used in the sigmoid. Top panels show distributions for $p_{mut}=0.05$ and bottom panels for $p_{mut}=1$. }  \label{fig:SIoneshotKin}
\end{figure}

Distributions of the probability of cooperating with an unknown agent in a one-shot game, are shown in Fig. \ref{fig:SIoneshot} and  Fig. \ref{fig:SIoneshotKin} for the non-kin and kin models, respectively.

\end{document}